\documentclass[12pt]{article}
\usepackage[a4paper,margin = 2cm]{geometry}
\usepackage[T1]{fontenc}
\usepackage{times}
\usepackage{amsthm}               
\usepackage{amsfonts}
\usepackage{setspace}
\usepackage{newtxtext,newtxmath}
\usepackage{graphicx}
\usepackage{rotating}
\usepackage{authblk}
\usepackage{xcolor}
\usepackage{natbib}
\usepackage{tikz}
\usepackage{bm}
\usepackage{bigints}
\usepackage{multirow}
\usepackage{float}
\usepackage{longtable}
\usepackage{lscape}
\usepackage{setspace}
\usepackage{mathtools}
\usepackage{multicol}
\usepackage{caption, subcaption}
\usepackage{pdfpages}
\usetikzlibrary{shapes,decorations,arrows,calc,arrows.meta,fit,arrows,shapes.arrows,
shapes.geometric,shapes.multipart,decorations.pathmorphing,positioning,shapes.swigs}
\tikzset{
    -Latex,auto,node distance =1 cm and 1 cm,semithick,
    state/.style ={ellipse, draw, minimum width = 0.7 cm},
    point/.style = {circle, draw, inner sep=0.04cm,fill,node contents={}},
    bidirected/.style={Latex-Latex,dashed},
    el/.style = {inner sep=2pt, align=left, sloped}
}

\newcommand{\T}{T}
\usepackage{bbm}

\newcommand{\p}{\mbox{P}}

\newcommand\independent{\protect\mathpalette{\protect\independenT}{\perp}}
\def\independenT#1#2{\mathrel{\rlap{$#1#2$}\mkern2mu{#1#2}}}
\newtheorem{assumption}{Assumption}
\newtheorem{proposition}{Proposition}
\newtheorem{theorem}{Theorem}

\title{Structural mean models for instrumented difference-in-differences}

\author[1]{Tat-Thang Vo}
\author[2]{Ting Ye}
\author[3]{Ashkan Ertefaie}
\author[1]{Samrat Roy}
\author[4]{James Flory}
\author[5]{Sean Hennessy}
\author[6]{Stijn Vansteelandt}
\author[1]{Dylan S. Small}
\affil[1]{Department of Statistics and Data Science, The Wharton School, University of Pennsylvania, USA}
\affil[2]{Department of Biostatistics, Hans Rosling Center for Population Health, University of Washington, USA}
\affil[3]{Department of Biostatistics and Computational Biology, University of Rochester, USA}
\affil[4]{Department of Subspecialty Medicine, Memorial Sloan Kettering Cancer Center, USA}
\affil[5]{Department of Biostatistics, Epidemiology and Informatics, Perelman School of Medicine, University of Pennsylvania, USA}
\affil[6]{Department of Applied Mathematics, Computer Science and Statistics, Ghent University, Belgium}

\begin{document}

\maketitle

\begin{abstract}
In the standard difference-in-differences research design, the parallel trends assumption may be violated when the relationship between the exposure trend and the outcome trend is confounded by unmeasured confounders. Progress can be made if there is an exogenous variable that (i) does not directly influence the change in outcome means (i.e. the outcome trend) except through influencing the change in exposure means (i.e. the exposure trend),  and (ii) is not related to the unmeasured exposure - outcome confounders on the trend scale. Such exogenous variable is called an instrument for difference-in-differences. For continuous outcomes that lend themselves to linear modelling, so-called instrumented difference-in-differences methods have been proposed. In this paper, we will suggest novel multiplicative structural mean models for instrumented difference-in-differences, which allow one to identify and estimate the average treatment effect on count and rare binary outcomes, in the whole population or among the treated, when a valid instrument for difference-in-differences is available. We discuss the identifiability of these models, then develop efficient semi-parametric estimation approaches that allow the use of flexible, data-adaptive or machine learning methods to estimate the nuisance parameters. We apply our proposal on health care data to investigate the risk of moderate to severe weight gain under sulfonylurea treatment compared to metformin treatment, among new users of antihyperglycemic drugs.
\end{abstract}

\section{Introduction}

The estimation of treatment effects in observational studies is often subject to bias due to unmeasured confounding. For instance, observational pharmacoepidemiological studies often utilize data from large administrative claim databases or electronic health records, which were not collected for research and may have incomplete/inaccurate information on potential confounding variables \citep{zhang2018addressing}.
In view of this concern, various analytical methods have been proposed to detect or control for unmeasured confounding \citep{uddin2016methods}. Among these approaches, instrumental variable and difference-in-differences  designs are very commonly used \citep{baiocchi2014instrumental,wing2018designing}. Instrumental variable methods makes use of an exogeneous variable that is associated with the exposure, but that does not directly affect the outcome and is independent of unmeasured confounders \citep{baiocchi2014instrumental}.
The difference-in-differences method is instead based on a comparison of the trends in outcome for two exposure groups, where one group consists of individuals who switch from being unexposed to exposed and the other group consists of individuals who are never exposed. Assuming that the outcomes in the two exposure
groups evolve in the same way over time in the absence of the exposure (i.e., the parallel trends assumption), the difference-in-differences method  is able to remove time-invariant bias caused by unmeasured confounders \citep{wing2018designing}. 

To further relax assumptions, the \textit{instrumented difference-in-differences} design has recently been proposed, which combines the strength of instrumental variable and difference-in-differences \citep{ye2020instrumented}. This method allows one to identify the treatment effects under a weaker set of assumptions than each parent method alone. As an example, in the standard difference-in-differences design, the parallel trends assumption could be violated when the relationship between the change in exposure mean (i.e. the exposure trend) and the change in outcome mean (i.e. the outcome trend) is confounded by unmeasured confounders. The instrumented difference-in-differences method overcomes this challenge by leveraging an exogenous variable that does not have any direct causal impact on the outcome trend except via the exposure trend, and is not associated with the unmeasured confounders on the trend scale \citep{ye2020instrumented}. Importantly, this so-called instrument for difference-in-differences need not itself be a valid instrumental variable for the considered exposure - outcome association, e.g. it can have a direct causal effect on the outcome that is not mediated through the exposure at each time point. 

In this paper, we aim to improve the utility of instrumented difference-in-differences by proposing structural mean models for this design. Structural mean models were first introduced by \citet{robins1994correcting} and \citet{robins1991correcting}, and then were extended to instrumental variable and other settings by \citet{vansteelandt2003causal}, \citet{hernan2006instruments}, \citet{tchetgen2010doubly}, among many others. Our contributions to this literature can be summarized as follows:

First, we propose a set of causal assumptions to identify the average exposure effect (by using the instrument for difference-in-differences) that is arguably weaker than the one previously proposed in \citet{ye2020instrumented}. We achieve this by considering novel additive structural mean models for instrumented difference-in-differences. One advantage of these structural mean models is that they allow the flexibility to model non-linear relations. Besides, when the proposed assumptions are violated, one can alternatively narrow the focus onto the additive average exposure effect among the exposed, which can avoid the assumption of no unmeasured exposure effect modifiers.

Thus far, the instrumented difference-in-differences method has only been developed for continuous outcomes. Our second contribution is to extend this method to a count outcome or a rare binary outcome. We achieve this by proposing multiplicative structural mean models for instrumented difference-in-differences. As in the additive case, under certain causal assumptions, the proposed multiplicative structural mean models allow one to identify and estimate the average treatment effect on the multiplicative scale, in the whole population or among the exposed, when a valid instrument for difference-in-differences is available.

Third, we develop robust and efficient estimation strategies for the parameters indexing the multiplicative structural mean models, using semi-parametric theory. Proposed estimators can achieve $\sqrt{n}$ rate of convergence to the parameters of interest, even when the nuisance functions are estimated at slower rates, e.g. by using flexible, data-adaptive or machine learning methods. We consider two different settings. In the first setting, the impact of the baseline covariate on the outcome in the structural mean models is characterized by some finite-dimensional parameter vector. In the second setting, it is left unspecified.  


\section{Additive structural mean models for instrumented difference-in-differences}

Assume that a random sample of a target population is followed up over two time points, i.e. $t=0$ and $t = 1$. For each individual $i$ in the sample, we observe $ O_i = (Z_i,  X_i, D_{0i}, Y_{0i}, D_{1i},Y_{1i})$; where $D_{ti}$ and $Y_{ti}$ are the respective exposure and outcome status observed at each time point $ t$ $(t=0,1)$, $ X_i$ is a vector of baseline covariates and $Z_i = 0,1$ is a binary instrument for difference-in-differences observed at baseline. The observations $(O_1, \ldots, O_n)$ are independent and identically distributed realizations of $ O = (Z, X,D_0, Y_0,D_1,Y_1)$. In Figure 1a, we describe the relationship between different variables by a causal diagram. 

Denote $Y_t^{d}$ the counterfactual outcome that would be observed at time point $t$ if the exposure $D_t$ were set to $d$ $(d=0,1)$. Throughout the rest of the paper, we will suppose that the following consistency assumption holds:
\begin{assumption}\label{consist} ~ $Y_t^d = Y_t$ when $D_t = d$, for all $t,d = 0,1$. 
\end{assumption}
\begin{figure}[h!!]
\centering
  \scalebox{0.7}{\subfloat[Original scale]{
  \begin{tikzpicture}[node distance =1.5 cm and 1 cm]
    \node[state,draw = none] (z) at (0,0) {$Z$};
    \node[state,draw = none] (d0) [right =of z] {$D_0$};
    \node[state,draw = none] (y0) [right =of d0] {$Y_0$};
    \node[state, draw = none] (d1) [right =of y0] {$D_1$};
    \node[state, draw = none] (y1) [right =of d1] {$Y_1$};
        
    \node[state,circle] (u0) [below =of y0] {$U_0$};
    \node[state,circle] (u1) [below =of y1] {$U_1$};

    \path (z) edge (d0);\path (d0) edge (y0);
    \path (y0) edge (d1);\path (d1) edge (y1);
    \path (u0) edge (y0);\path (u0) edge (d0); \path (u0) edge (u1);\path (u0) edge (d1); \path (u0) edge (y1);
    \path (u1) edge (y1);\path (u1) edge (d1);    
    \path (z) edge (u0);  
    \draw [->] (z) to [out=35, in =145] (y0);
    \draw [->] (z) to [out=35, in =145] (y1);
    \draw [->] (d0) to [out=35, in =145] (d1);
    \draw [->] (d0) to [out=35, in =145] (y1);
    \draw [->] (y0) to [out=35, in =145] (y1);
\end{tikzpicture}
} \quad
  \subfloat[Trend scale]{
  \begin{tikzpicture}[node distance =1.5 cm and 1 cm]
    \node[state,draw = none] (z) at (0,0) {$Z$};
    \node[state,draw = none] (d) [right =of z] {$\Delta D$};
    \node[state,draw = none] (y) [right =of d] {$\Delta Y$};
    
    \node[state,circle] (u) [below =of y] {$U_0, U_1$};
    \path (z) edge (d);\path (d) edge (y);
    \path (u) edge (y);\path (u) edge (d); 
\end{tikzpicture}}}
\caption{Data generating mechanism. The baseline covariates $X$ are omitted to simplify the figure.}
\end{figure}
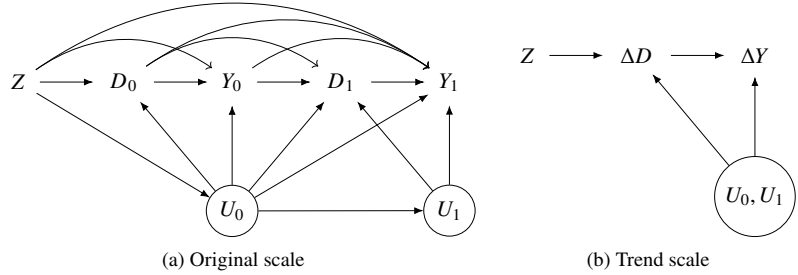
Our first aim is to make inferences about the conditional average exposure effect given $ X$ on the additive scale, assuming that this effect is unchanged over time, e.g. when the study period only spans a short period of time. Denote $\beta(x)$ the average exposure effect given $ X = x$, one then has $
\beta( x) = E(Y^1_{t} - Y^0_{t}| X = x)$ for $t=0,1$.
When $X$ is empty or when $ X$ does not include any effect modifier, $\beta( x) = \beta$ expresses the average exposure effect.

Under causal diagram 1a, $Z$ cannot be used as a standard instrumental variable to estimate $\beta(x) $. For instance, the exclusion restriction assumption is violated because $Z$ may have a direct effect on $Y$ that is not via $D$. Similarly, a standard difference-in-differences analysis is invalid because of the relationship between the exposure trend $\Delta D = D_1 - D_0$ and the outcome trend $\Delta Y = Y_1 - Y_0$ being confounded by unmeasured variables $U = (U_0,U_1)$ (figure 1b). Progress can however be made if conditional on $ X$, the variable $Z$ does not have any direct effect on the outcome trend except via the exposure trend, and moreover $Z$ is independent of the unmeasured exposure-outcome confounders on the trend scale \citep{ye2020instrumented}. Such exogenous variable $Z$ is referred to as an instrument for difference-in-differences, and will allow one to estimate $\beta(x)$ by a Wald-type estimator derived from the identity:
\begin{align}\label{wald}
\beta(x) = \frac{E(Y_1 - Y_0\mid X= x, Z=1) - E(Y_1 - Y_0\mid X= x, Z = 0)}{E(D_1 - D_0\mid X= x, Z=1) - E(D_1 - D_0\mid X= x, Z = 0)}
\end{align}
provided that $E(D_1 - D_0\mid X = x, Z=1) \ne E(D_1 - D_0\mid X = x, Z = 0)$.
In the discussion below, we will show that this identification result can be obtained by viewing $\beta$ as a parameter indexing a particular additive structual mean model for instrumented difference-in-differences. The advantage of such model is that it easily enables extensions, e.g. when the treatment effect on the multiplicative scale is of more interest. An additive structual mean model can be formally expressed as:
\begin{align}\label{smm1}
E(Y_1^{d^*} - Y_0^{d}\mid X =  x, Z) =  \beta( x) \times (d^* - d) + m(x), 
\end{align}
for all $d,d^*$, where $\beta(x)$ and $m(x)$ are unknown. 
This model embodies the assumptions that (i) the average outcome trend given $ X$ under the same exposure over time is unchanged across strata defined by $Z$, i.e $E(Y_1^{d} - Y_0^{d}\mid X, Z=1) = E(Y_1^{d} - Y_0^{d}\mid X, Z=0) = m( X) $, and that (ii) the (time-independent) average treatment effect is constant across stratum defined by $Z$, i.e. $E(Y_t^{d^*} - Y_t^{d}\mid X, Z=1) = E(Y_t^{d^*} - Y_t^{d}\mid X, Z=0) = \beta(X)(d^* - d)$. In other words, $Z$ is assumed to not modify the effect of time and of exposure on the outcome, which was previously referred to as the \textit{independence \& exclusion restriction} assumption by \citet{ye2020instrumented}. To link $\beta (x) $ to the observed data, they further assume that there are no unmeasured confounders of the relationship between $Z$ and $(D_t, Y_t)$ that simultaneously modify the effect of $Z$ on $D_t$ and of $Z$ on $Y_t$, given $X$ \citep{ye2020instrumented}. Here, we alternatively consider the following sequential ignorability assumption:
\begin{assumption}[Sequential ignorability]
\label{u0u1} 
There exists $(U_0, U_1)$ possibly unmeasured such that (i) $Y_0^d \independent D_0\mid U_0, X, Z$ and $Y_1^d \independent D_1\mid U_0,U_1,Z, X, Y_0, D_0$ for $d = 0,1$, and (ii) $U_0$ and $U_1$ do not modify the exposure effect on the additive scale.
\end{assumption}
Assumption \ref{u0u1} is arguably more intuitive. The first component 2(i) is quite standard and commonly assumed in other settings with longitudinal data such as mediation analysis \citep{imai2010general} or repeatedly measured exposure and outcome \citep{hernan2001marginal}. The second component 2(ii) essentially assumes that all exposure effect modifiers on the additive scale are measured. This is a strong assumption that will be relaxed below. As an example, model (\ref{smm1}) and  Assumption \ref{u0u1} hold when the outcome generating mechanism at each time point obeys the following linear models:
\begin{align*}
&Y_1 = \alpha_1 + \beta_1 D_1 + \beta_2 D_1 X +\gamma_0U_0 + \gamma_1U_1 + \delta Z + \epsilon_1,\\
&Y_0 = \alpha_0 + \beta_1 D_0 + \beta_2 D_0 X +\gamma_0U_0 + \delta Z + \epsilon_0,
\end{align*}
where $\epsilon_1$ and $\epsilon_0$ are mean-zero, normally distributed random errors (conditional on the variables in these respective models). In the Supplementary Material, we show that under Assumption \ref{u0u1}, $\beta(X)$ can be linked to the observed data by identification result (\ref{wald}).

To avoid the assumption that all effect modifiers are measured, one could alternatively focus on the conditional average exposure effect among the exposed. When $X = x$, this effect can be expressed as:
$\beta^* (x) = E(Y_t^1 - Y_t^0\mid D_t = 1, X = x)$ for $t=0,1
$, 
which is also assumed to be unchanged over time. To estimate $\beta^*(x)$, consider the following model: 
\begin{align}\label{eq8}
E(Y_t^d\mid D_t = d, X = x, Z) - E(Y_t^0\mid D_t = d,X = x, Z) = \beta^*(x)\times d~~~~~\mathrm{for}~~~~d=0,1
\end{align}
which embodies the assumption that $Z$ itself does not modify the exposure effect among the exposed (given $ X$). Under the additional assumption that $Z$ does modify the effect of time on the outcome (given $ X$) when the whole population is unexposed, i.e.
\begin{assumption} \label{as.trend}
~~~$Y_1^{0} - Y_0^{0} \independent Z \mid X,$ 
\end{assumption} 
the parameter $\beta^*(x) $ can be expressed as the right-hand side of expression (\ref{wald}). These combined assumptions are arguably weaker than the independence and exclusion restriction assumptions embodied in model (3), and moreover do not require exposure effect modifiers to be fully measured at both time points.
\section{Multiplicative structural mean models for instrumented difference-in-differences}
\subsection{Identification}
In this section, we extend the above discussion to a multiplicative structural mean model for instrumented difference-in-differences. The aim is to identify and estimate $\beta(x)$ defined on the multiplicative scale, still assuming that such effect is unchanged over time, i.e.
$\beta(x) = E(Y_t^{1}\mid X = x)/E(Y_t^{0}\mid X = x)$ for $t=0,1$.
To achieve this, consider the following model:
\begin{align}\label{smm2}
E(Y_1^{d^*}\mid X = x, Z) = E(Y_0^{d}\mid X = x, Z)e^{\beta( x) \times (d^* - d) + m( x)}.
\end{align}
Model (\ref{smm2}) can be viewed as an extension of model (\ref{smm1}) to the multiplicative scale. This model embodies the assumption that $Z$ does not modify the effect of time on the outcome on the multiplicative scale, i.e.
$E(Y_1^{d}\mid X, Z=z)/E(Y_0^{d}\mid X, Z=z) = e^{m( X)}$ for $z=0,1$,
nor the effect of the exposure on the outcome on the multiplicative scale, i.e. $E(Y_t^{d^*}\mid X, Z=z)/E(Y_t^{d}\mid X, Z=z) = e^{\beta(X) \times (d^* - d)}$ for $z=0,1$. In the Supplementary Material, we prove that under the sequential ignorability Assumption \ref{u0u1}, $\beta ( X) $ is linked to the observed data by the following moment condition:
\begin{align}\label{moment.rr}
E\{Y_1 e^{-\beta(X) D_1} - Y_0 e^{-\beta(X)D_0 + m(X)} \mid X, Z\} = 0
\end{align}
As for the additive structural mean model, one can alternatively target the conditional average exposure effect among the exposed (on the multiplicative scale) to avoid the assumption of no unmeasured effect modifiers 2(ii). As for the additive setting, one needs to impose the alternative assumptions that (a) the conditional average exposure effect among the exposed is unchanged over time, (b) $Z$ does not modify the effect of the exposure among the exposed and (c)  $Z$ does not modify the effect of time on the outcome among the unexposed. In the Supplementary Material, we further discuss this extension and show that under the aforementioned assumptions, the conditional average exposure effect among the exposed on the multiplicative scale can also be linked to the observed data by an expression similar to (\ref{moment.rr}).

\subsection{Estimation without baseline covariates}
We first discuss the setting in which no adjustment for baseline covariates is needed. In model (\ref{smm2}), one then has $X$ empty, $m(X) = 0$ and $\beta(x) = \beta$ quantifying the average exposure effect. The moment condition (\ref{moment.rr}) then implies that:
\begin{align}\label{est.eq.rr}
E(Y_1e^{-\beta D_1}\mid Z = 1)~ E(Y_0e^{-\beta D_0}\mid  Z = 0) = E(Y_0e^{-\beta D_0}\mid Z = 1)~ E(Y_1e^{-\beta D_1}\mid Z = 0)
\end{align}
Solving the sample analog of this equation returns a consistent estimator $\hat\beta$ for $\beta$. Obtaining a closed-form expression for $\hat\beta$ is not possible in general cases. However, when the exposure is binary ($D_0, D_1 = 0,1$), equation (\ref{est.eq.rr}) can be rewritten in a quadratic form as:
\begin{align*}
(E_{111}E_{000} - E_{110}E_{001})\theta^2 - (E_{11}E_{000} + E_{111}E_{00} - E_{10}E_{001} - E_{110}E_{01})\theta + E_{11}E_{00} - E_{10}E_{01} = 0
\end{align*}
where $\theta = e^{-\beta}- 1$, $E_{tz} = E(Y_t\mid Z = z)$ and $E_{ttz} = E(Y_tD_t\mid Z = z)$ for $t,z =0,1$. 

The asymptotic distribution of $\hat\beta$ can be established using standard $M$-estimation theory.

\subsection{Estimation with baseline covariates}
We now discuss estimation strategies when the set of baseline covariates $X$ is non-empty. For this, we will assume that $ \beta( x) = \beta_0 + \beta_1^\T x$, but the proposed methods will work for any other finite-dimensional parametrization of $\beta( x)$. With a slight abuse of notation, we denote $\beta^\T = \begin{pmatrix} \beta_0 & \beta_1^\T\end{pmatrix}$ as the $k$-dimensional vector of parameters indexing $ \beta(x)$. 

We consider two settings. In the first setting, we let the covariate function $m(X)$ in the structural mean model (\ref{smm2}) be correctly parametrized, in the sense that $m(X) = m(X, \gamma)$ for some finite-dimensional parameter $ \gamma$. In the second setting, $m( X)$ is unspecified. In both cases, we will denote $\epsilon = \epsilon(O,\beta, m(\cdot)) = Y_1e^{-\beta( X) D_1} - Y_0 e^{-\beta( X) D_0 + m( X)}$. The moment condition (\ref{moment.rr}) implies that $E(\epsilon\mid X, Z) = 0$. 

\textbf{Setting 1: $m( X)$ specified}. 
Assume that $m( X, \gamma)$ is correctly specified. To construct consistent estimators for $\beta$ and $\gamma$, we first note that these parameters actually index a semi-parametric model  $\mathcal{M}$, represented by the class of distributions $\mathcal{P}$ of the observed data satisfying (\ref{moment.rr}), i.e. for which $\int \epsilon(o,\beta, \gamma)d\mathcal{P}(d_0,y_0,d_1,y_1\mid X, Z,\beta,\gamma) = 0$. From this restriction, one can derive the space of all influence functions (i.e. the orthogonal nuisance tangent space) of $\mathcal{M}$. Because of the deep connection between (asymptotically linear) estimators for a given model and the influence functions under that model, if we can find all the influence functions for $\mathcal{M}$, we can characterize all regular asymptotic linear estimators for $\theta = \begin{pmatrix} \beta^\T & \gamma^\T \end{pmatrix}^\T$ up to asymptotic equivalence \citep{tsiatis2006semiparametric,kennedy2016semiparametric}.

\begin{theorem}\label{theo1}
Suppose that Assumption \ref{consist}, \ref{u0u1} and model (\ref{smm2}) hold. The space of all influence functions for $\theta = \begin{pmatrix} \beta^\T & \gamma^\T \end{pmatrix}^\T$ under the proposed specification of $m( X, \gamma)$ in model (\ref{smm2}) is $\Lambda_1^{\perp} = \big\{ d^{q\times1}( X, Z)\cdot \epsilon \big\}$,
where $q$ denotes the dimension of $ \theta$ and $ d^{q\times1}( X, Z)$ is an arbitrary $q$-dimensional vector function of $ X$ and $Z$ that satisfies
\[ E\bigg\{ d^{q\times1}( X, Z) \bigg(\frac{\partial \epsilon}{\partial \theta}\bigg)^\T\bigg\} = I^{q\times q}.\]
Here and below, $I^{q\times q}$ denotes the $q\times q$ identity matrix.
\end{theorem}

Theorem \ref{theo1} suggests that $\theta$ can be estimated by solving the sample equivalent of the moment condition $E\{ d( X,Z)\epsilon\} =  0$, where $ d( X,Z)$ is an arbitrary non-trivial $q$-dimensional vector function of $ X$ and $Z$, e.g. $ d^\T( X,Z) = \begin{pmatrix} 1 & X^\T & Z\end{pmatrix}$. A straightforward application of the $M$-estimation method then allows one to derive the asymptotic variance of ${\hat\theta}$ obtained from this approach. More precisely, $\sqrt{n}({\theta} - \theta_0)$ converges in distribution to:
\[ N\bigg[ 0,~ E\bigg\{-\frac{\partial f}{\partial \theta}( O,\theta_0)\bigg\}~ \mathrm{var}\big\{ f( O,\theta_0)\big\} ~ E\bigg\{-\frac{\partial f}{\partial \theta} ( O,\theta_0)\bigg\}^{-1,\T} \bigg] \]
where $\theta_0$ denotes the true values of $\theta$ and $ f( O, \theta_0) = d( X,Z) \epsilon$ denotes the estimating function. 

Alternatively, one can obtain the asymptotic variance of ${\hat\theta}$ by using non-parametric bootstrap sampling.

For completeness, we also derive in the Supplementary Material the efficient influence function among the elements of $\Lambda_1^\perp$, by projecting the score of $\theta$ (under the true parametric submodel) on $\Lambda_1^\perp$. However, we do not recommend the use of this efficient influence function in practice. First, it involves $\mathrm{var}(\epsilon\mid X, Z)^{-1} = E^{-1}(\epsilon^2\mid X, Z)$ as a nuisance parameter. The efficiency of the resulting estimator is thus local in the sense that it is only attained when this variance can be estimated consistently at sufficiently fast rates. Even when a proper estimate can be obtained for $\mathrm{var}(\epsilon\mid X, Z)$, the inverse of this variance can make the resulting estimator for $\theta$ become very unstable, which makes it difficult to perform well in practice. 

\textbf{Setting 2: $m(X)$ unspecified}. 
We now discuss a more general setting where the function $m( X)$ in model (\ref{smm2}) is left unspecified. For this, consider first an easier case where $m( X, Z)$ is a priori known. By a similar proof as in Theorem \ref{theo1}, one can show that under Assumption \ref{consist} and \ref{u0u1}, the orthogonal complement of the nuisance tangent space in model (\ref{smm2}) (given that it is correctly specified) is $\Lambda_1^{\perp} = \{d^{k\times 1}(X, Z)\epsilon\}$, where  $d^{k\times 1}( X, Z)$ is arbitrary but satisfies: \[ E\bigg\{ d^{k\times1}( X, Z) \bigg(\frac{\partial \epsilon}{\partial \theta}\bigg)^\T\bigg\} = I^{k\times k}.\] 
To recognize that $m( X)$ is unknown, we then need to determine the subspace of mean-zero functions in $\Lambda_1^{\perp}$ that is additionally orthogonal to the nuisance scores for $m(X)$. 

\begin{theorem}\label{theo3}
Suppose that Assumptions \ref{consist}, \ref{u0u1} and model (\ref{smm2}) hold. The orthogonal complement of the nuisance tangent space of model (\ref{smm2}) when $m(X)$ is left unspecified is $\Lambda_2^{\perp} = \big\{\big[ d^{k\times 1}( X, Z) -  d^{*,k\times 1}( X, Z)
\big]\epsilon \big\},$
where $ d^{k\times 1}( X, Z)$ is an arbitrary $k$-dimensional function of $ X$ and $Z$ that satisfies \[ E\bigg\{\big[ d^{k\times 1}( X, Z) - d^{*,k\times 1}( X,Z)
\big] \frac{\partial \epsilon}{\partial \beta}\bigg\} = 1\]
with: 
\[ d^{*,k\times 1}(X,Z) = \frac{\lambda( X, Z)\sigma^{-2}( X, Z)}{E\big\{\lambda^2( X, Z)\sigma^{-2}( X, Z)\mid X\big\}} E\big\{ d^{k\times 1}( X, Z)\lambda( X, Z)\mid X\big\}\]
and $\lambda( X, Z) = E(Y_0e^{-\beta( X) D_0}|Z, X)/E(Y_0e^{-\beta( X) D_0}| X).$
\end{theorem}
A direct consequence of Theorem \ref{theo3} is that elements in $\Lambda_2^{\perp}$ have mean zero even when $m( X)$ is mispecified, i.e. $m( X) \ne m_0( X)$, where $m_0( X)$ denotes the true form of $m( X)$ that is unknown, provided that $\sigma^{-2}( X,Z)$, $\lambda(X,Z)$, $E\{d(X,Z)\lambda(X,Z)\mid X\}$ and $E(\lambda^2( X, Z)\sigma^{-2}( X, Z)\mid X)$ are correctly specified. Note that postulating parametric models for these nuisance parameters is not entirely satisfactory, as it may easily lead to model mispecification and incompatibility. Besides, the estimating functions in $\Lambda_2^{\perp}$ are highly complex (e.g. due to the presence of many complicated nuisance parameters), which may lead to convergence issues in practice. 

To remedy this, consider the element $\nu$ of $\Lambda_2^\perp$ corresponding to $d^{k\times 1}(X,Z) = g^{k\times 1}( X, Z) - E\big\{ g^{k\times 1}( X, Z)\lambda( X, Z)\mid X\big\}$, where $g^{k\times 1}( X, Z)$ is an arbitrary $k$-dimensional vector function of $X$ and $Z$ satisfying the conditions in Theorem \ref{theo3}. This element $\nu$ can be alternatively expressed as:
\[\nu= \big\{ g( X, Z) - E\big( g( X, Z)\lambda( X, Z)\mid X\big)\big\}\big\{Y_{1}e^{-\beta( X) D_{1}} - Y_0e^{-\beta(X) D_{0} + m(X)}\big\},\]
Theorem 2 then implies that $E(\nu) = 0$ when $m(X) \ne m_0(X)$, given that only $\lambda(X,Z)$ and $E\big( g( X, Z)\lambda( X, Z)\mid X\big)$ are consistently estimated. In our current setting with binary exposure, by fixing $m( X) = 0$, the moment condition $E(\nu) = 0$ can be reexpressed as:
\begin{align}\label{eq.nonpara}
 E\bigg[  \underbrace{\bigg\{ d( X, Z) - \frac{(e^{- \beta( X)}-1)  a_1( X) +  a_2( X)}{(e^{- \beta( X)}-1) a_3( X) +  a_4( X)}\bigg\}}_\text{$A(O)$} \underbrace{\bigg\{(Y_1D_1 - Y_0D_0)(e^{-\beta( X)}-1) + Y_1 - Y_0
 \bigg\}}_\text{$B(O)$} \bigg]
 = 0
 \end{align}
Here, we denote 
 $a_1(X)  = E\{Y_{0}D_{0} g( X,Z)\mid X\}$,
 $ a_2( X)  = E\{Y_0 g( X,Z)\mid X\}$, 
 $a_3( X) = E\{Y_{0}D_{0}\mid X\}$ and 
 $a_4(X) = E\{Y_{0}\mid X\}$. 

We now construct an estimation strategy for the parameter vector $\beta^{k\times 1}$ based on the moment condition (\ref{eq.nonpara}). Since $ \beta^{k\times 1}$ is the zero of this moment condition, it can be viewed as a well defined model-free population parameter without reference to the original model (\ref{smm2}). This suggests that one can work under the non-parametric model and
estimate $\beta^{k\times 1}$ by using semi-parametric theory, to enable fast $\sqrt{n}$ rates of convergence to the parameter of interest. This is achievable even when nuisance functions $ a_1( X)$ to $a_4( X)$ are estimated at slower rates, e.g. using flexible data-adaptive or machine learning methods. 

In what follows, we will focus on the special case where $\beta( X) = \beta$ (i.e. $ X$ are not effect modifiers and $k=1$) for the sake of simplicity. In the Supplementary Material, we generalize the discussion to more general cases where $\beta( X)$ is a non-constant parametric function of $ X$ (i.e.  $k>1$).
Denote $\theta = e^{-\beta} - 1$. To characterize its influence function under the non-parametric model (so as to obtain a non-parametric estimator), we first rewrite $\theta$ as $\theta = \theta(\mathcal{P})$ to stress that $\theta$ is a functional of the observed data distribution $\mathcal{P}$. In what follows, we perturb $\theta$ in the direction $\tilde{\mathcal{P}}_t$ of a point mass at single observation $\tilde{o}$ of $ O$, i.e. $\tilde{\mathcal{P}}_t = (1-t)\mathcal{P} + t\mathbbm{1}_{\tilde{o}}$ where $\mathbbm{1}_{\tilde{o}}$ denotes the Dirac delta function at $\tilde{\bm o}$. The efficient influence function of $\theta$ at observation $\tilde{ o}$ under the non-parametric model for the observed data can then be identified by evaluating the Gateaux derivative of $\theta(\mathcal{P}_t)$ with respect to $t$ at $t = 0$, that is 
\[\phi( \tilde{o}, \theta,\eta) = \frac{d \theta(\mathcal{P}_t)}{dt}\bigg |_{t=0}\]
where $\eta = c(a_1,a_2,a_3,a_4,a_5,a_6)$ is the vector of all nuisance parameters. For a more detailed guidance on influence functions, see \citet{hines2022demystifying} and \citet{kennedy2016semiparametric}.
\begin{proposition}\label{theo4}
Under certain regularity conditions, it can be shown that the influence function of $\theta$ under the non-parametric model is:
\begin{align*}
 &\phi(\bm O, \theta,\eta) = - C^{-1}A( O) B( O) + C^{-1} \times \\
 &\times \bigg[ \bigg\{ \frac{\theta Y_0g(X,Z)(D_0+1)- a_1\theta-a_2}{\theta a_3 + a_4} - \frac{(\theta a_1 + a_2)(\theta Y_0D_0 - \theta a_3 + Y_0 - a_4)}{(\theta a_3 + a_4)^2}\bigg\} E\big\{B( O)\mid X \big\}\bigg] 
 \end{align*}
 where $C=C( O,\theta,\eta) =  E\big\{(a_2a_3-a_1a_4)(\theta a_3 + a_4)^{-2}B( O) + A( O)(Y_1D_1-Y_0D_0)\big\}$.
\end{proposition}
As $E\{\phi( O, \theta, \eta)\} = 0$ by construction, one can obtain an estimate $\hat\theta$ by solving the sample analog of this equation, $
\sum_i^n \phi( O_i,\hat\theta,\hat{ \eta}) = 0$, 
where $\hat{\eta} = (\hat a_1, \hat a_2, \hat a_3, \hat a_4, \hat a_5, \hat a_6)$ denotes an estimate for $\eta$, possibly obtained by flexible data-adaptive or machine learning methods. Assume that $\hat{\eta}$ converges in probability to some $\eta_1$ that might be potentially different from the true value of the nuisance parameter $ \eta$. In the Supplementary Material, we prove that the remainder term: 
$R( \eta,\eta_1) := \theta(\eta_1) - \theta(\eta) + E\{\phi(O,\eta_1)\}$
is a second order term involving only products of the type $E[ c(\eta,\eta_1)\{f(\eta_1)-f(\eta)\}\{g(\eta_1)-g(\eta)\}]$. This result will be useful when establishing the asymptotic properties of $\hat\theta$, as shown in the theorem below.
\begin{theorem}\label{theta.var}
(Asymptotic normality and efficiency) Assume that (i) the second-order term $R({\hat\eta}, \eta)$ is $o_P(n^{-1/2})$ and (ii) the class of functions $\{ \phi(\eta,\theta'): |\theta'-\theta| < \delta, ||\eta - \eta_1|| < \delta\}$ is Donsker for some $\delta>0$ and such that $\mathrm{pr}\{\phi(\eta,\theta') - \phi(\eta_1,\theta)\}^2 \rightarrow 0$ as $(\eta,\theta') \rightarrow (\eta_1, \theta)$, then: $\hat\theta(\hat{\eta}) = \theta( \eta) 
+ \frac{1}{n} \sum_{i=1}^n \phi(O_i, \eta) + o_P(1), $
due to which $\sqrt{n}(\hat\theta_1 - \theta)$ converges in distribution to $N(0,\zeta^2)$, where $\zeta^2 = \mathrm{var}\{ \phi(O,\eta)\}$ is the non-parametric efficiency bound.
\end{theorem}
The proof of this theorem follows the general proof presented in \citet{chernozhukov2017double}. Some remarks are noteworthy here. First, condition (i) for asymptotic normality in Theorem \ref{theta.var} requires that all components of $\hat{\eta}$ converges in $L_2(P)$ norm to their true counterparts in $\eta$ at faster than $n^{1/4}$-rate, to ensure that the remainder term $R(\hat\eta,\eta)$ is of second order (see above). Under certain conditions, this can be satisfied by data-adaptive algorithms such as regression trees \citep{wager2015adaptive}, neural networks \citep{chen1999improved}, and highly adaptive lasso \citep{van2017generally}. 

Condition (ii) (i.e. Donsker condition) restricts the flexibility of the nuisance estimators, but Donsker classes still cover many complex functions such as Lipschitz functions and so forth \citep{van2000asymptotic,kennedy2016semiparametric}. Alternatively, one can avoid condition (ii) by using cross-fitting in the estimation procedure, given that the nuisance estimators $\hat\eta$ are consistent and satisfy condition (i). Let $V_1, \ldots , V_Q$
denote a random partition of the index set $\{1, \ldots , n\}$ into $Q$ sets of approximately similar size. For each index $q$, the training sample is given by $T_q = \{1, \ldots, n\} \setminus V_q$. Let $\hat{\eta}^q$ denote the estimator of $\eta$, obtained by training the corresponding prediction algorithm using only data in the sample $T_q$. Further, let $q_i$ denote the index of the validation set which contains observation $i$. The proposed estimator may be adapted to cross-fitting by substituting all occurrences of $\hat{\eta}( O_i)$ by $\hat{\eta}^{q_i}( O_i)$ in the estimation procedure.

A direct consequence of Theorem \ref{theta.var} is that the (asymptotic) behavior of the proposed estimator $\hat\theta$ is the same as if the nuisance parameters $\eta$ were known. As such, one can quite easily obtain a sandwich estimator of the asymptotic variance of $\hat\theta$ as the sample variance of $\phi(O, \hat{ \eta})$. This variance estimate may be used to construct Wald-type confidence intervals. 
\section{A simulation study}
In this section, we conduct a simulation study to assess the finite sample performance of the proposed approaches. The aim of the analysis is to estimate the average causal effect of a binary exposure on a count outcome (setting 1 and 3), or on a rare binary outcome with frequency around $10-12\%$ (setting 2 and 4). Assume that each patient is followed up over two time points (i.e. longitudinal data structure). At each time point, the exposure-outcome relationship is confounded by an unmeasured variable $U_t$ ($t=0,1$). In settings 3 and 4, adjusting for a baseline covariate $X$ is needed for the binary instrument for difference-in-differences $Z$ to be valid. The average treatment effect is $\beta(X) = 0$ across all settings. Other details about the data generating mechanism are provided in table 1.
\begin{center}
\begin{table}[htb!!]
\centering
\caption{Simulation study: Data generating mechanism}
\scalebox{0.8}{\begin{tabular}{ p{1.5cm} p{3cm} p{13cm}}
\textbf{Setting} & \textbf{Characteristics} & \textbf{Data generating mechanism }\\
1 & Baseline &
 $Z \sim B(N,0.5)$
 \\
 & Time $t=0$ &
$U_0 \sim N(0.5, 1)$ \\
& & $E(D_0 \mid U_0, Z) = \mathrm{expit}(1 - Z + U_0)$ \\
& & $E(Y_0\mid Z, U_0, D_0) = \exp(-1+ 0D_0 + 0.5U_0 + 0.5Z)$
\\
 & Time $t = 1$ & 
$U_1 \sim N(0.5, 1)$ \\
& & $E(D_1\mid U_1, U_1, Z) = \mathrm{expit}(-1 + Y_0 + U_1 + Z)$ \\
& & $E(Y_1\mid Z, U_1, D_1, Y_0) = \exp(-1 + 0 D_1 + 0.5U_1 + 0.5Z)$ 
\\
\\
2 & Baseline & $Z \sim B(N,0.5)$
\\
& Time $t=0$ & $U_0 \sim U(0, 1)$ \\
& & $E(D_0\mid X, U_0, Z) = \mathrm{expit}(-0.85 - Z + U_0)$\\
& & $E(Y_0\mid X, Z, U_0, D_0) = \exp(-3.7+ 0 D_0 + U_0 + Z) $
\\
& Time $t = 1$ & $U_1 \sim U(0, 1)$ \\
& & $E(D_1\mid X, U_0, U_1, Z) = \mathrm{expit}(0.272 + Y_0 + U_1 + Z)$\\
& & $E(Y_1\mid X, Z, U_1, D_1, Y_0) = \exp(-3.9 + 0 D_1 + U_1 + Z)$ 
\\
\\
3 & Baseline &
 $X = \min\{P(0.5) + 0.5, 2.5 \}$\\
 & & $E(Z\mid X) = \mathrm{expit}(-0.5 + X)$
 \\
 & Time $t=0$ &
$U_0 \sim N(0.5, 1)$ \\
& & $E(D_0\mid  U_0, Z) = \mathrm{expit}(1 - Z + U_0 + X_0)$ \\
& & $E(Y_0\mid Z, U_0, D_0) = \exp[-1+ 0D_0 + 0.5U_0 + 0.5Z + 0.25X + 0.15\sin(X)]$
\\
 & Time $t = 1$ & 
$U_1 \sim N(0.5, 1)$ \\
& & $E(D_1\mid U_1, U_1, Z) = \mathrm{expit}(-1 + Z + U_1 + Y_0 + X)$ \\
& & $E(Y_1\mid Z, U_1, D_1, Y_0) = \exp[-1 + 0 D_1 + 0.5U_1 + 0.5Z + 0.35X + 1.70\sin(X)]$ 
\\
\\
4 & Baseline & $X = \min\{P(0.5) + 0.5, 3.5 \}$\\
 & & $E(Z\mid X) = \mathrm{expit}(-0.8 + X)$
\\
& Time $t=0$ & $U_0 \sim U(0, 1)$ \\
& & $E(D_0\mid X, U_0, Z) = \mathrm{expit}(-0.85 - Z + U_0 + X)$\\
& & $E(Y_0\mid X, Z, U_0, D_0) = \exp[-1.8+ 0 D_0 - 1.5U_0 - 0.25Z + 0.15X + 0.15\sin(X)] $
\\
& Time $t = 1$ & $U_1 \sim U(0, 1)$ \\
& & $E(D_1\mid X, U_0, U_1, Z) = \mathrm{expit}(0.272 + Y_0 + 0.5U_1 + 0.5Z + 0.5X)$\\
& & $E(Y_1\mid X, Z, U_1, D_1, Y_0) = \exp[-3 + 0 D_1 - 1.5U_1 - 0.25Z + 0.35X + 1.70\sin(X)]$ 
\\
\end{tabular}}
\end{table}
\end{center}

Across all settings, using $Z$ as a standard instrument will return a biased estimate for $\beta$ as  the exclusion restriction assumption is violated (i.e. $Z$ has a direct effect on $Y_t$ that does not go through $D_t$). The instrumented difference-in-differences approach is alternatively used to analyze the data as follow:

In settings 1 and 2 with no observed covariates, we estimate $\beta$ by solving equation (\ref{est.eq.rr}).

In settings 3 and 4, the function $m(X)$ in the underlying structural mean model (\ref{smm2}) has the form $m(X) = \beta_0 + \beta_1X + \beta_2 \mathrm{sin}(X)$. We consider three approaches to estimate $\beta$.
\textbf{Approach (A1)}: $m(X)$ is mis-specified as $m(X) = \delta_0 + \delta_1 X$. The parameter vector $\theta = \begin{pmatrix} \delta_0 & \beta & \delta \end{pmatrix}^\T$ is estimated by solving the equation $\sum_{i=1}^N d(X,Z) \epsilon = 0$, where $ d(X,Z) = \begin{pmatrix} 1 & X & Z\end{pmatrix}^\T$.
\textbf{Approach (A2)}: $m(X)$ is correctly specified as shown above. The parameter vector $\theta = \begin{pmatrix}  \beta_0 & \beta & \beta_1 & \beta_2 \end{pmatrix}^\T$ is estimated by solving the equation $\sum_{i=1}^N  d(X,Z) \epsilon =  0$, where $ d(X,Z) = \begin{pmatrix} 1 & X & Z & \mathrm{sin}(X) \end{pmatrix}^\T$. 
\textbf{Approach (A3)}: $m(X)$ is unspecified. $\beta$ is estimated by the non-parametric approach discussed in section 3.3. The nuisance parameters involved in this approach are estimated by using the super learner algorithm \citep{van2007super}, whose library includes the main terms generalized linear model, the multivariate adaptive regression splines and the highly adaptive lasso. Although cross fitting is required to ensure valid inference for approach A3 without relying on the Donsker condition, we will not consider it here to shorten the computational time of the simulation study.

Three sample sizes, $n=\{5, 10, 15\}\times 10^3$, are considered in each setting. In setting 4 (rare binary outcome with observed baseline covariates), two other sample sizes of $n=\{20, 25\}\times 10^3$ are additionally considered to further evaluate the asymptotic properties of the proposed approaches. In each setting, we assess (i) the $\sqrt{n}$-consistency of the obtained estimator $\hat\beta$ for $\beta$, (ii) the ratio between the variance estimate of $\hat\beta$ and the true variance of $\hat\beta$ (calculated across all simulations), and (iii) the coverage of the 95\% Wald confidence interval for $\beta$. We implement $10^3$ simulations in each setting.

Results of this simulation study are visualized in figures 2 and 3. Numerical data to reproduce these figures are also provided in the Supplementary Material. 
When $X$ is empty and $m(X)=0$ (settings 1 and 2), the proposed method returns a valid estimate $\hat\beta$ for $\beta$ that is $\sqrt{n}$-consistent (figure 2a). In setting 2 (rare binary outcome), the variance of $\hat\beta$ is slightly underestimated when the sample size is small (figure 2b). This results in a (slight) over-coverage of the 95\% CI (figure 2c). Such a problem, however, disappears when the sample size is sufficiently large ($n=15\times 10^3$).

When $X$ is non-empty and $m(X)\ne 0$ (settings 3 and 4), the estimation approach based on Theorem 1 only provides a valid estimate for $\beta$ (i.e. $\sqrt{n}$-consistent) when the function $m(X)$ is correctly specified (i.e. approach A2). In contrast, the non-parametric approach A3 can obtain a valid estimate for $\beta$ without having to specify $m(X)$ (figure 3a).
When the outcome is a rare binary variable (setting 4), the performance of both approaches can be worsened if the sample size is insufficiently large (figure 3b-c).

\begin{figure}
    \centering
	\subfloat[Root-N Bias]{\includegraphics[scale=0.4]{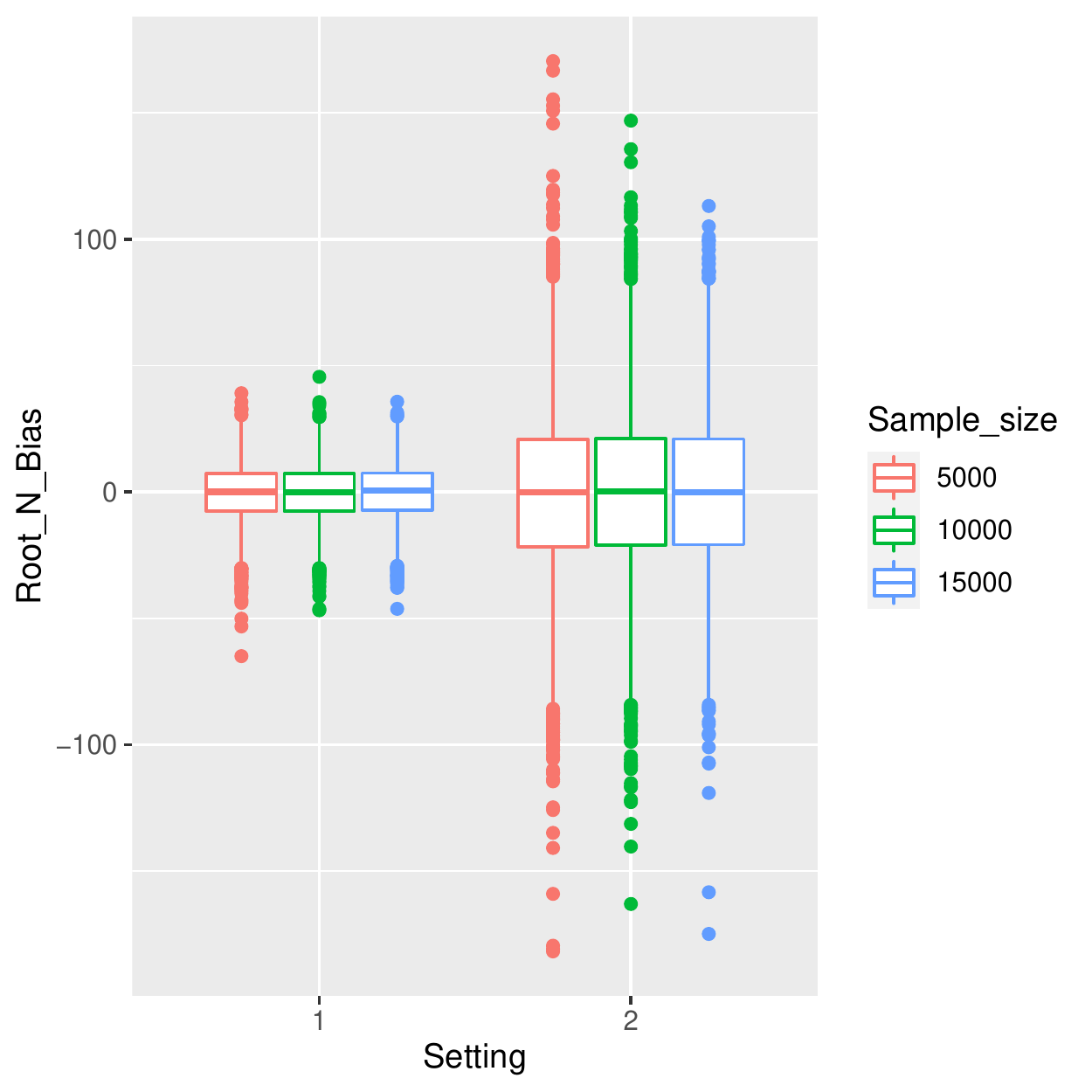}}
	\subfloat[Variance Ratio]{\includegraphics[scale=0.4]{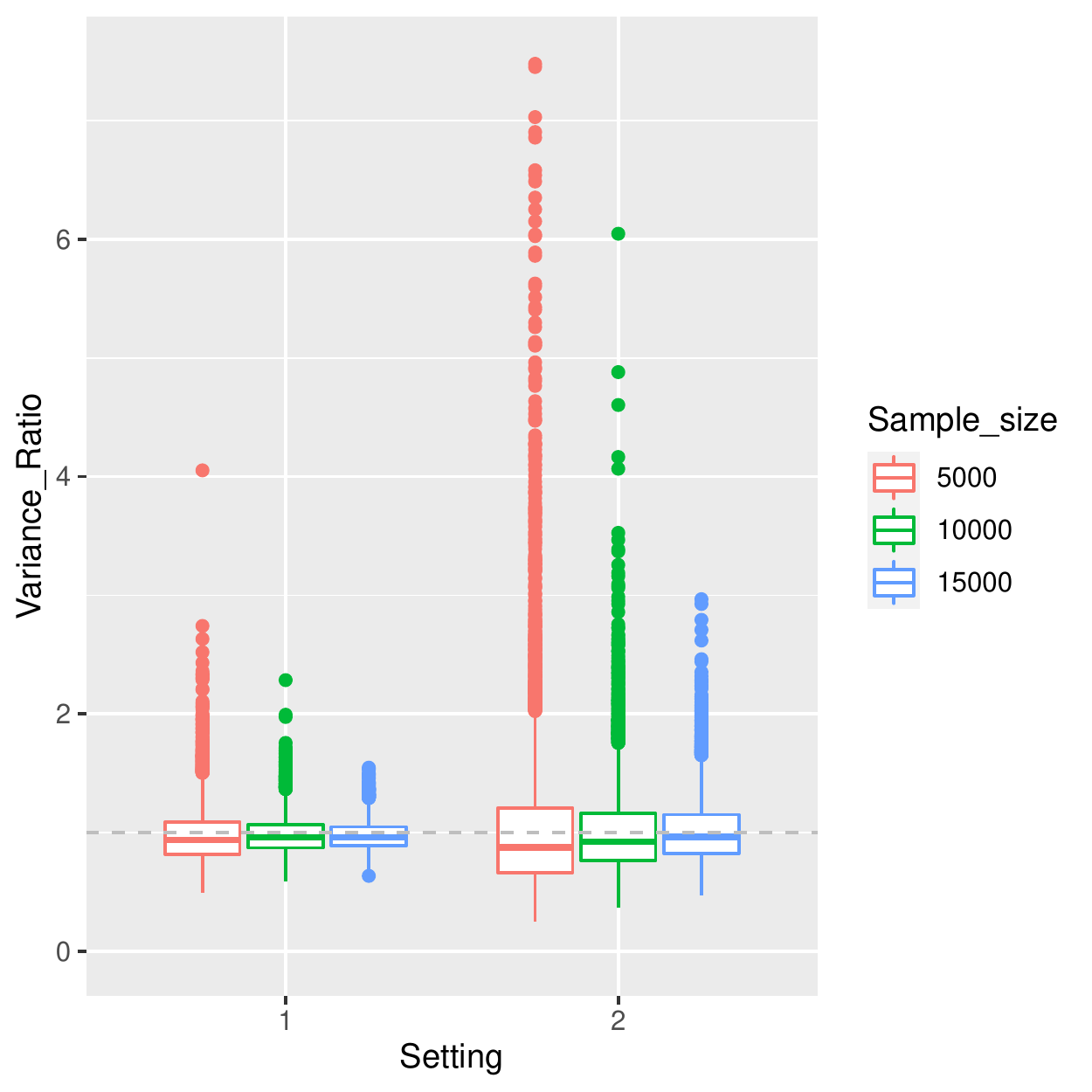}}
	\subfloat[Coverage]{\includegraphics[scale=0.4]{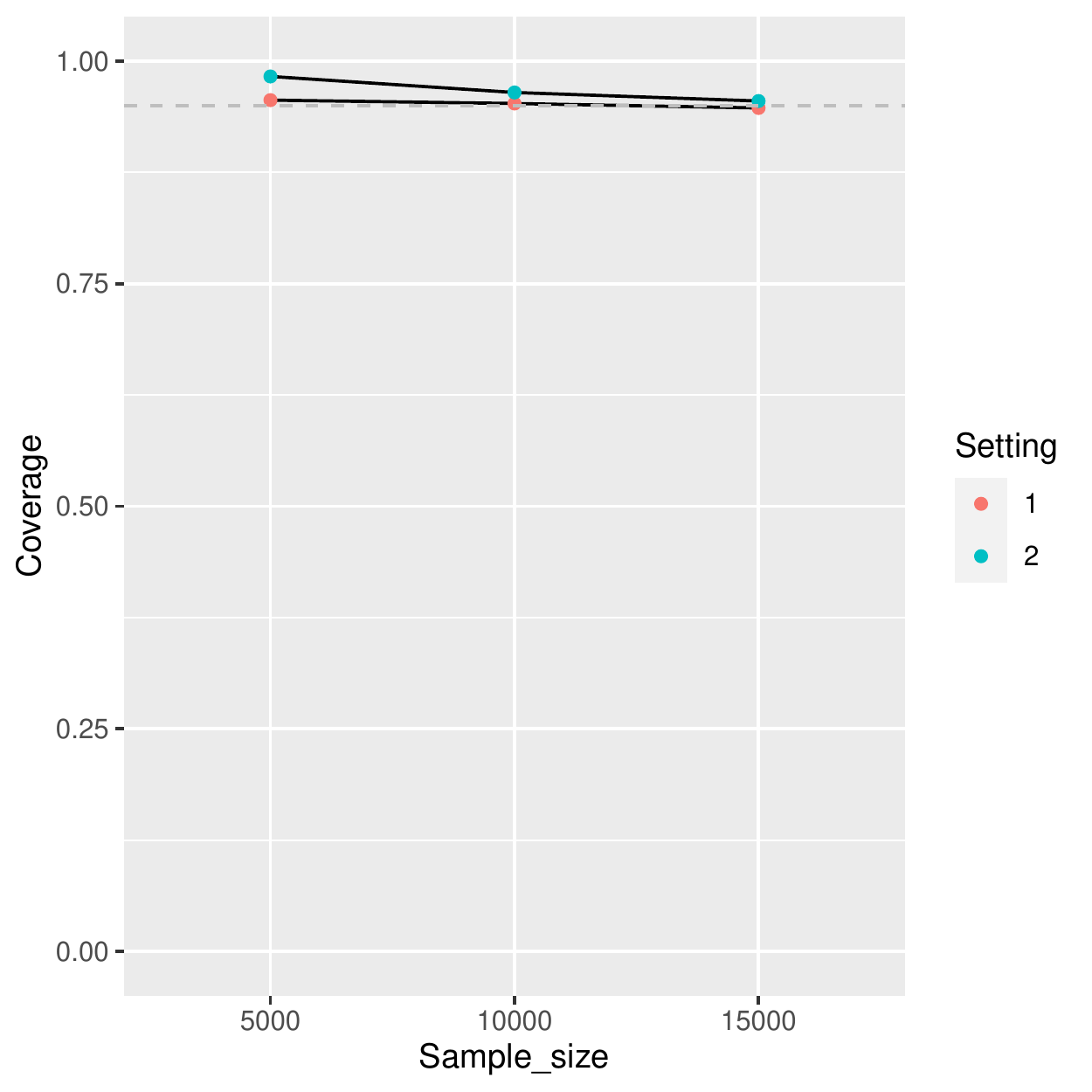}}
    \caption{Simulation results - Setting 1 and 2. (a): the distribution of $\sqrt{n}$-bias, i.e. $\sqrt{n}(\hat\beta_i - \beta)$, where $\hat\beta_i$ denotes the estimate of $\beta$ obtained from simulation $i$; 
    (b): the distribution of the ratio between the variance estimate $\hat V(\hat\beta_i)$ and the true variance $V(\hat\beta)$; (c): coverage of the 95\% Wald confidence interval for $\beta$}
    \label{fig2}%
\end{figure}

\begin{figure}
    \centering
    \subfloat[S3. Root-N Bias]{\includegraphics[scale=0.4]{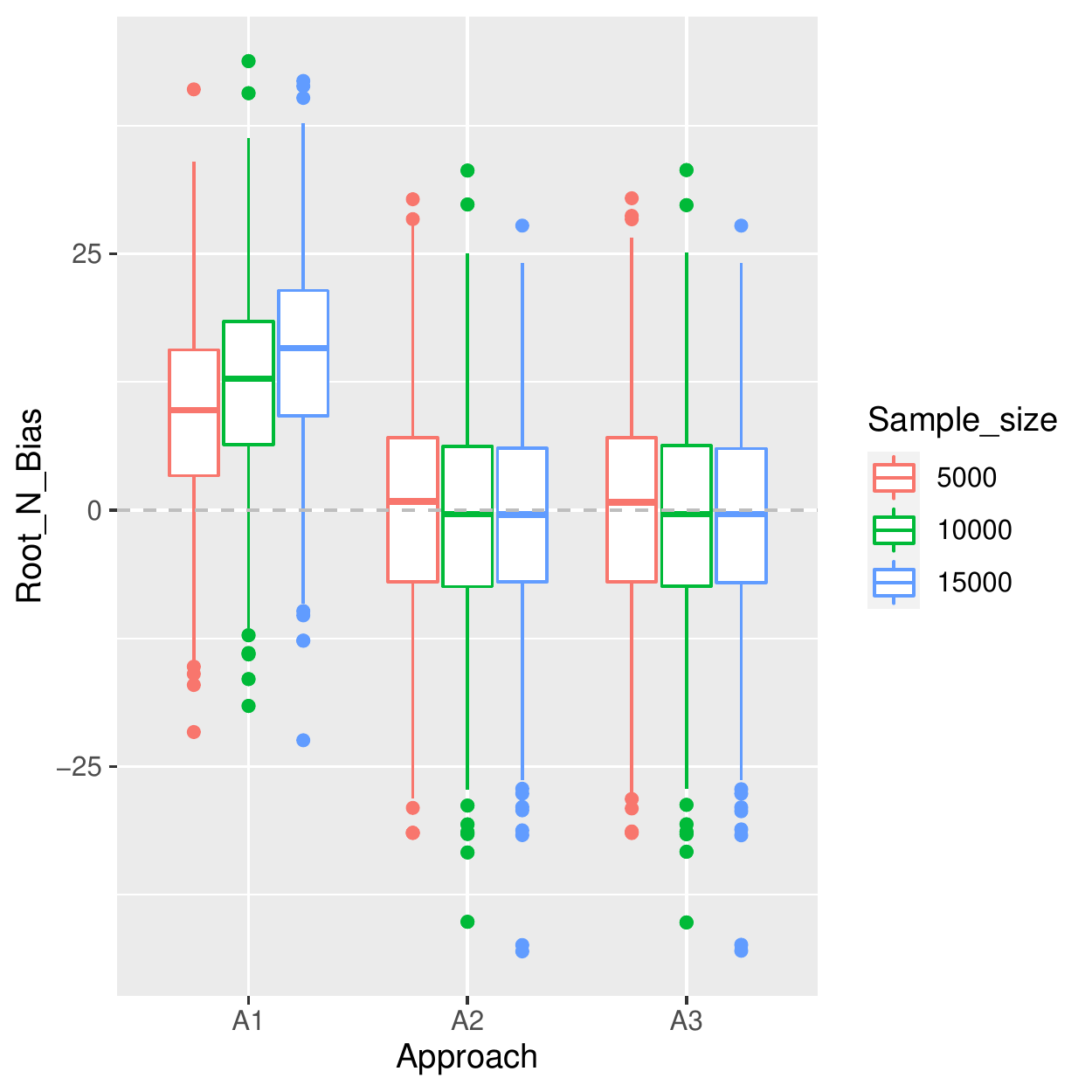}}%
    \subfloat[S3. Variance Ratio]{\includegraphics[scale=0.4]{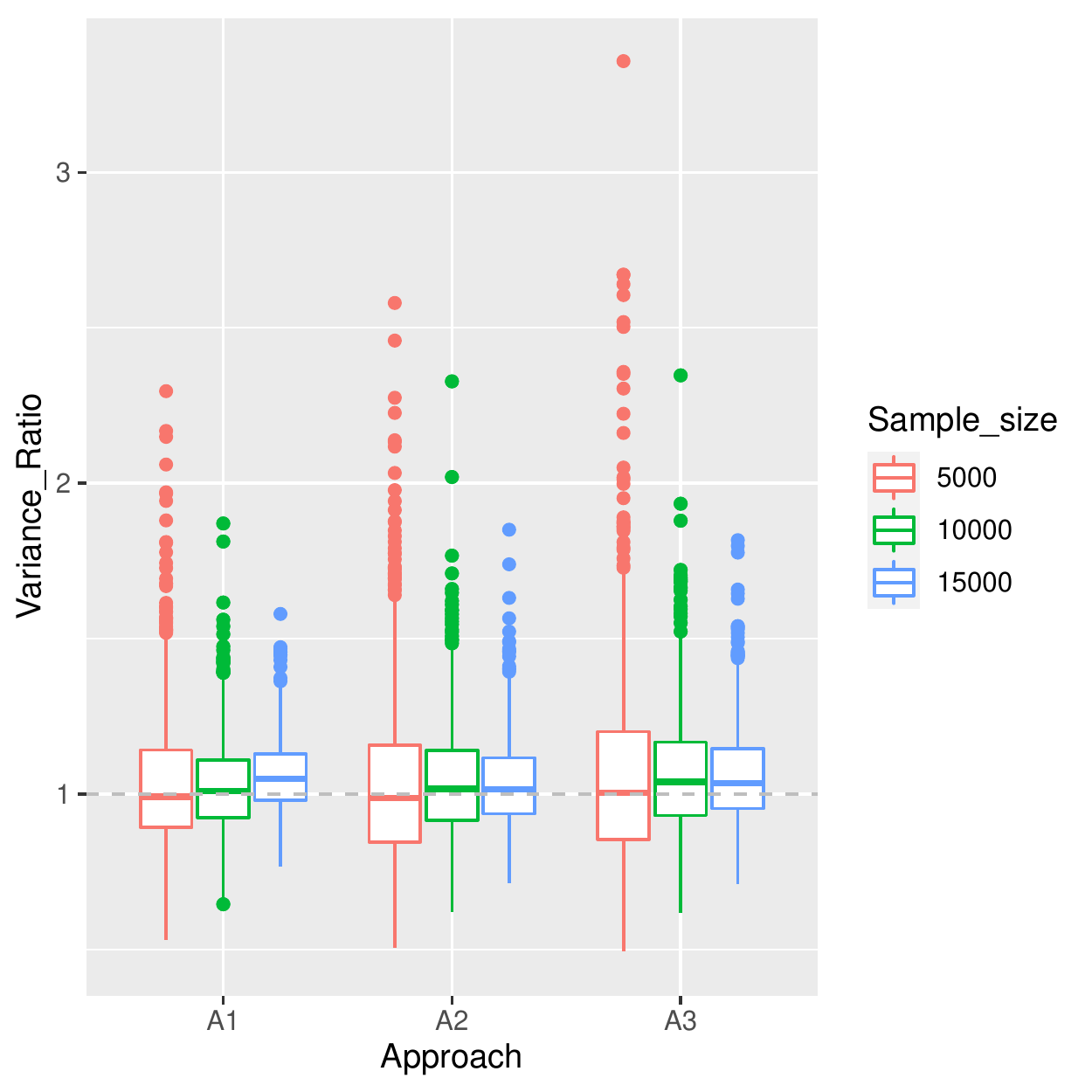}}%
    \subfloat[S3. Coverage]{\includegraphics[scale=0.4]{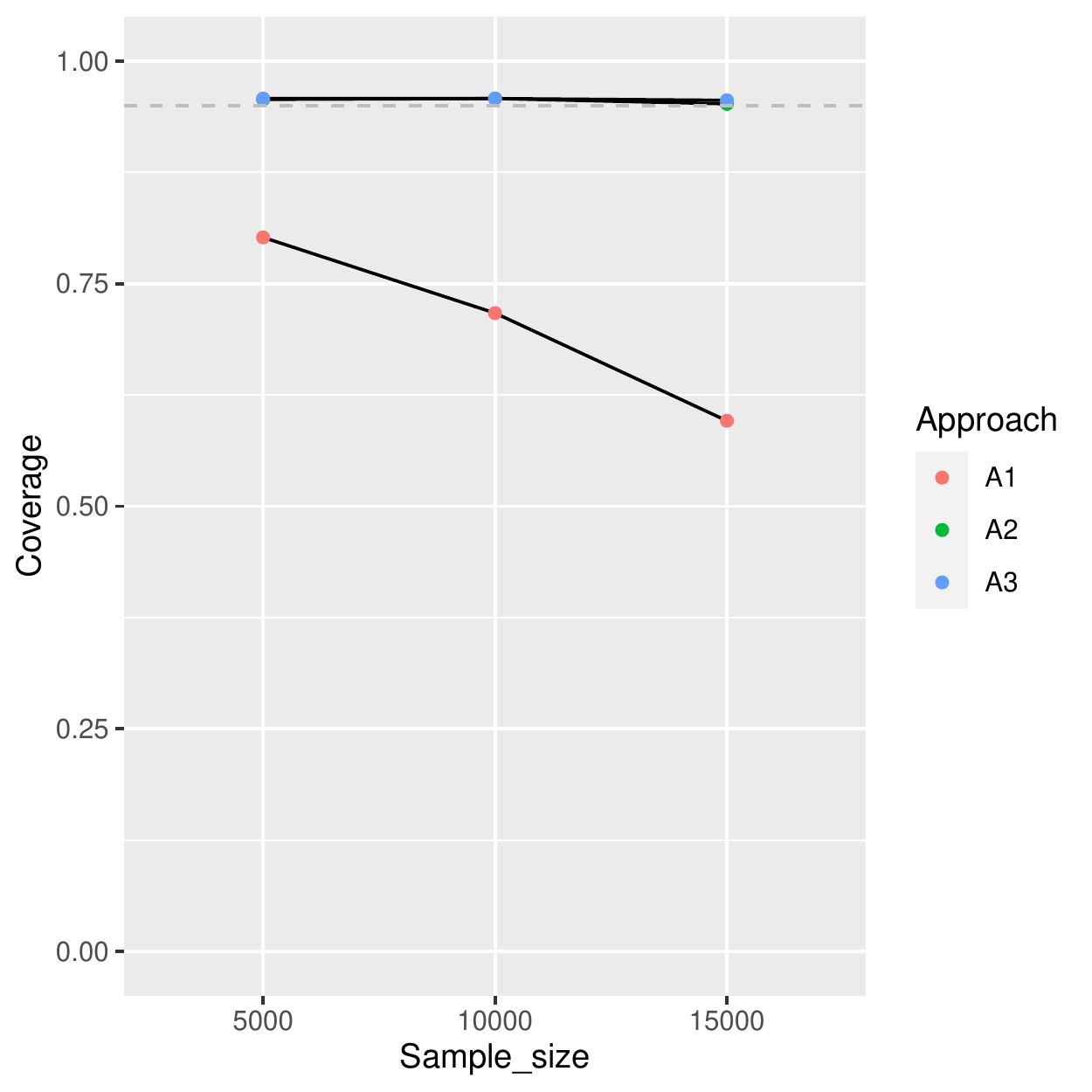}}%
    
    \subfloat[S4. Root-N Bias]{\includegraphics[scale=0.4]{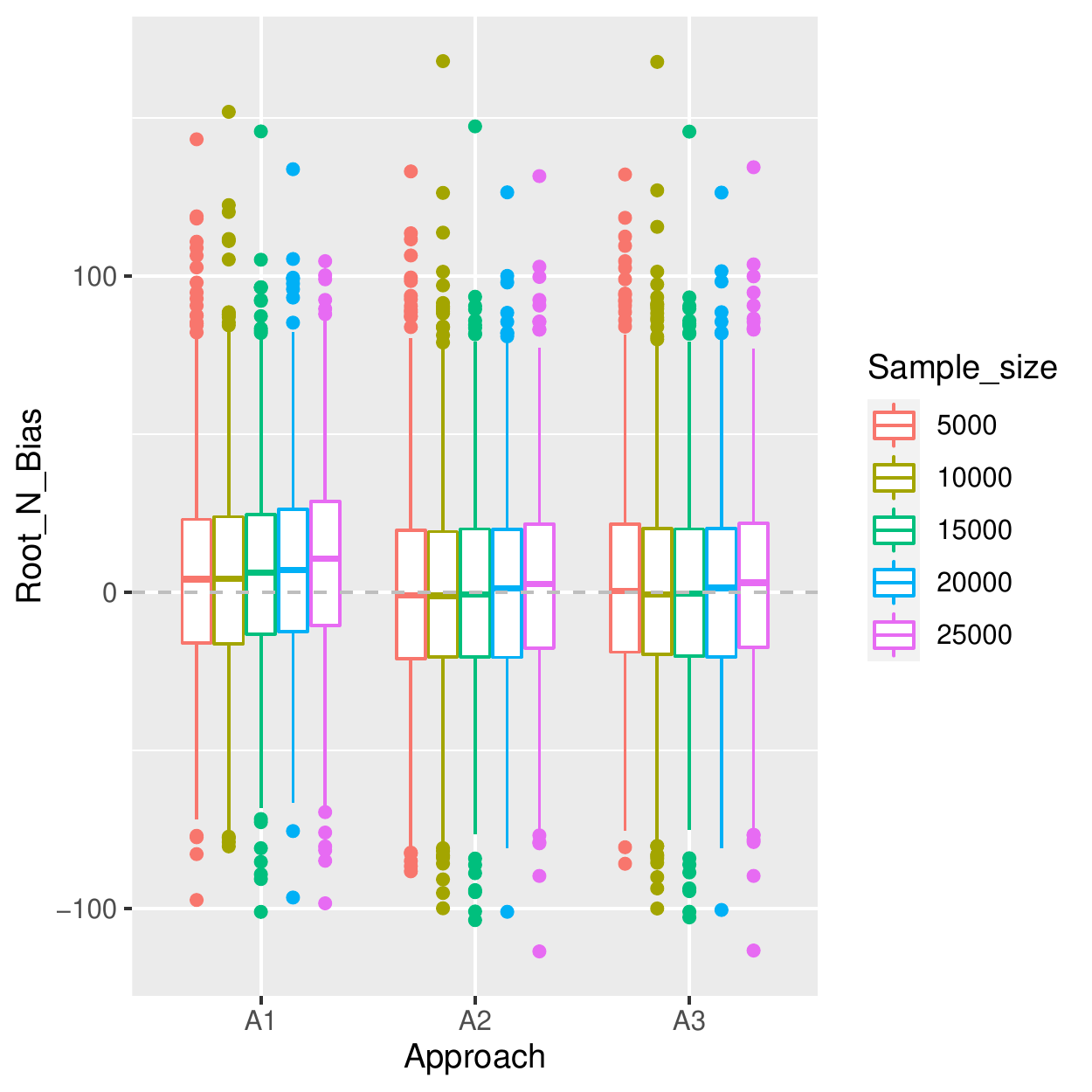}}%
    \subfloat[S4. Variance Ratio]{\includegraphics[scale=0.4]{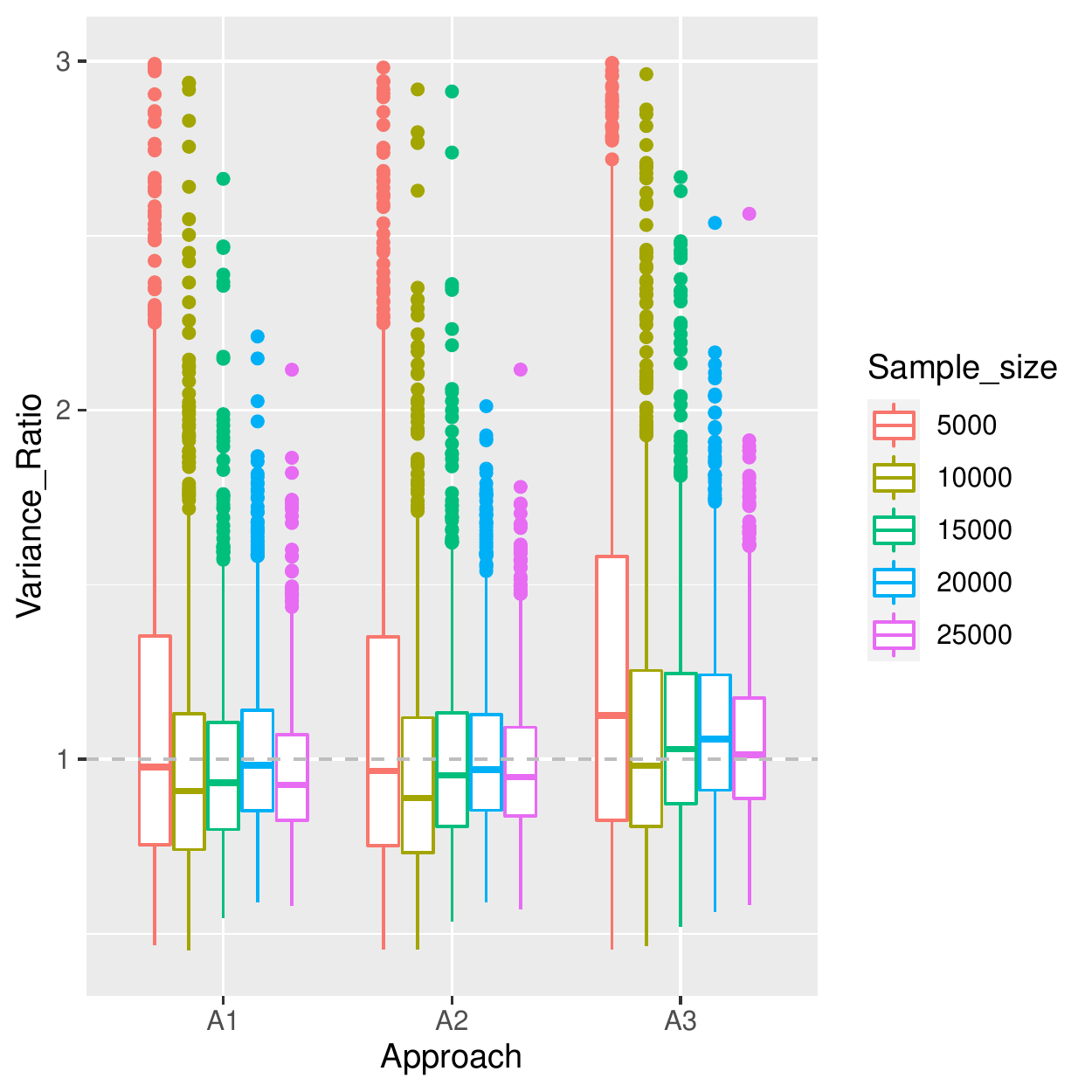}}%
    \subfloat[S4. Coverage]{\includegraphics[scale=0.4]{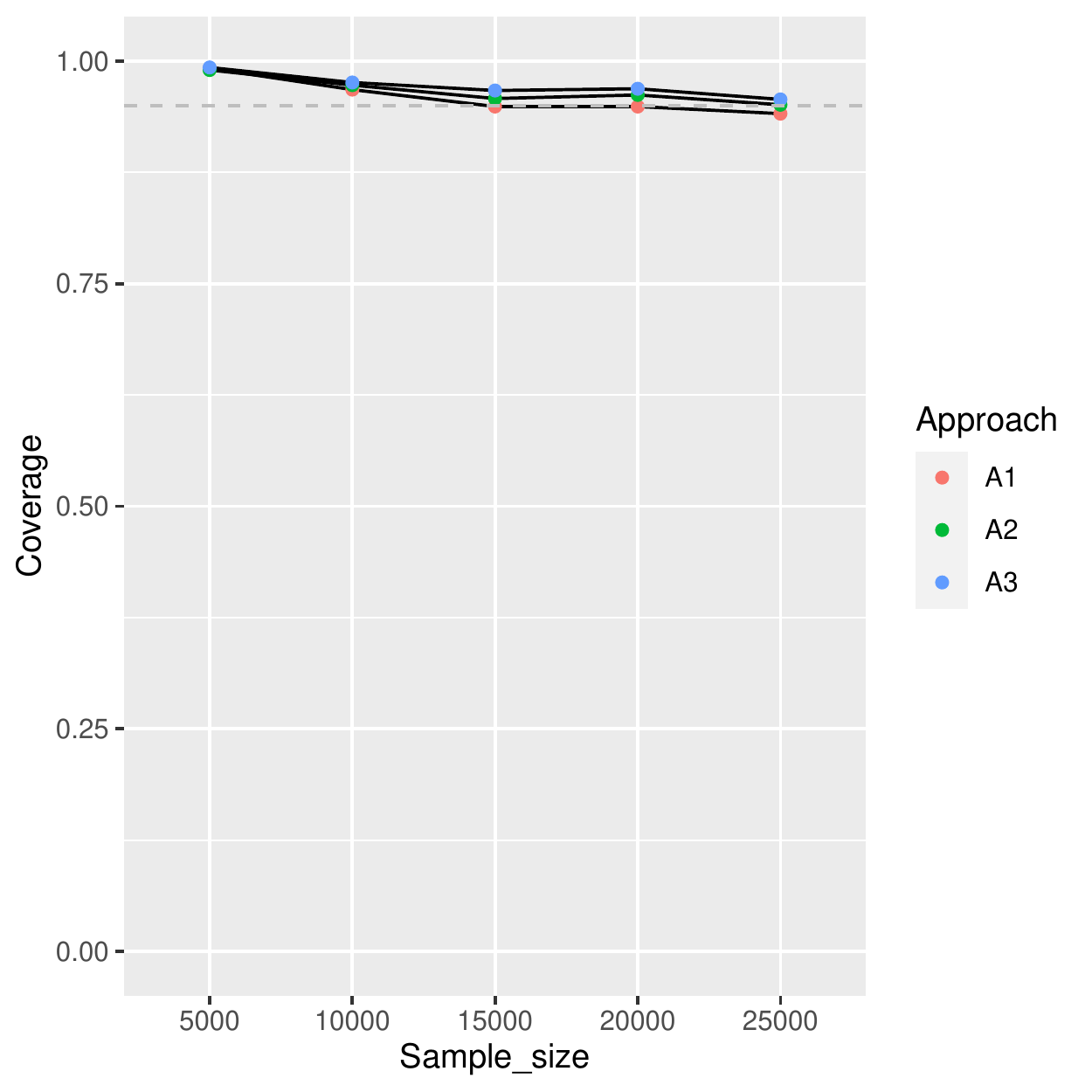}}%
    \caption{Simulation results - Setting 3 (S3) and setting 4 (S4). (a) and (d): the distribution of $\sqrt{n}$-bias; (b) and (e): the distribution of the ratio between the variance estimate and the true variance of $\beta$. (c) and (f): coverage of the 95\% Wald confidence interval for $\beta$ }
    \label{fig3}%
\end{figure}
\section{Extension to repeated cross-sectional data structure}
Thus far, we have discussed the instrumented difference-in-differences method for longitudinal or panel data, in which each individual is followed-up over two time points. In this section, we extend the above results to the repeated cross-sectional, or “pseudo-longitudinal”  data structure \citep{pan2022repeated}. In this setting, $D_t$ and $Y_t$ are evaluated on an independent sample at each time point $t$. For instance, respondents of an annual survey in one year are different from those in the prior year. It is thus commonly assumed that there is no overlap in the samples between different periods \citep{pan2022repeated}. 
To formalize this, denote $ O^* = (Z, X, D, Y, T)$ the observed data of each individual in a repeated cross-sectional study, where $T=0,1$ denotes the time point, $Y = Y_1I(T = 1) + Y_0I(T=0)$ and $D = D_1I(T = 1) + D_0I(T=0)$. For every stratum defined by levels of $Z$ and $X$, the collected data at each time point is a random sample from the population of interest, that is:
\begin{assumption}\label{missing} ~ $T \independent (Y_1, Y_0, D_1, D_0) \mid X, Z$
\end{assumption}

When $ X$ is empty, $m(X) = 0$ and $\beta( X) = \beta$, Assumption \ref{missing} implies that:
\begin{align}\label{eq12}
E(Ye^{-\beta D}\mid T=1,Z = 1) = E(Ye^{-\beta D}\mid T=0,Z = 1) ~\frac{E(Ye^{-\beta D}\mid T=1,Z = 0)}{E(Ye^{-\beta D}\mid T=0,Z = 0)}
\end{align}
Solving the sample analog of this equation will return a consistent estimator $\hat\beta$ for $\beta$. A simple application of the Delta method allows one to establish the asymptotic properties of $\hat\beta$.

Consider now the structural mean model (\ref{smm2}) with $X$ being non-empty and $m( X) = m( X, \gamma)$ correctly parameterized by some finite-dimensional parameter $\gamma$. To identify the orthogonal nuisance tangent space $\Lambda_3^\perp$ of model (\ref{smm2}) under the repeated cross-sectional data structure $ O^*$, one need to map the elements in $\Lambda_1^\perp$ to those in $\Lambda_3^\perp$. For this, note that for every $U_{ O^*} \in \Lambda_3^\perp$, the mean of $U_{ O^*}$ given $O$ (calculated with respect to the true distribution $\mathcal{P}$ of the full data structure $O$) must equal some element $U_{O} \in \Lambda_1^\perp$. The same remark also allows one to establish the orthogonal nuisance tangent space $\Lambda_4^\perp$ of model (\ref{smm2}) in the repeated cross-sectional setting, when $m( X)$ is left unspecified.

\begin{theorem}\label{theo41}
Suppose that Assumptions \ref{consist}, \ref{u0u1}, \ref{missing} and model (\ref{smm2}) hold. When the observed data structure is $ O^* = (Z, X, D, Y, T)$, the orthogonal complement of the nuisance tangent space of model (\ref{smm2}) under the parametrization $m(X) = m( X,\gamma)$ is:
\[\Lambda_3^{\perp} = \big\{ d^{q\times1}( X, Z)\cdot \pi( O,\theta) + s^{q\times 1}( X,Z) \cdot \big[T - \p(T = 1\mid Z, X)\big]\big\},\] 
where
\[ \pi ( O,\theta) = \frac{TYe^{-\beta(X) D}}{\p(T = 1|Z,X)} - \frac{(1 - T)Ye^{-\beta( X) D+ m( X)}}{1 - \p(T = 1\mid Z, X)}\]
and $ d^{q\times1}( X, Z)$ and $ s^{q\times1}( X, Z)$ are arbitrary $q$-dimensional vector functions of $ X$ and $Z$ that satisfies 
\[ E\bigg\{ d^{q\times1}( X, Z) \bigg(\frac{\partial \pi( O,  \theta)}{\partial \theta}\bigg)^\T\bigg\} = I^{q\times q}.\]
Here, $q$ denotes the dimension of the parameter vector $\theta = \begin{pmatrix} \beta^\T & \gamma^\T
\end{pmatrix}^\T$.
In contrast, the orthogonal complement of the nuisance tangent space of model (\ref{smm2}) under the data structure $O^*$, when $m( X)$ is left unspecified is:
\[\Lambda_4^{\perp} = \big\{ \big[ d^{k\times 1}( X, Z) -  d^{*,k\times 1}( X, Z)\big]\cdot \pi( O, \beta) + q^{k\times 1}( X,Z) \cdot \big[T - \p(T=1\mid Z, X)\big]\big\},\]
where $ d^{k\times 1}( X, Z)$ and $ q^{k\times 1}(Z, X)$ are arbitrary $k$-dimensional functions of $X$ and $Z$ that satisfy: 
\[E\bigg\{\big[ d^{k\times 1}( X, Z) -  d^{*,k\times 1}( X, Z)\big] \frac{\partial\pi( O, \beta)}{\partial \beta}\bigg\} =  I^{k\times k}\] 
and $ d^{*,k\times 1}( X,Z)$ is defined as in theorem 2, but with: \[\lambda( X, Z) = \frac{E\big[(1-T)Ye^{-\beta( X) D}\mid Z,  X\big]\big / \p(T=0\mid  X, Z)}{E\big[(1-T)Ye^{-\beta( X) D}\mid X\big]/\p(T=0\mid X)}.\]
\end{theorem}
A consequence of Theorem \ref{theo41} is that to estimate the parameters indexing model (\ref{smm2}), one needs to additionally model the nuisance parameter $P(T=1| X, Z)$. The estimation strategies that we have previously discussed in section 3 can then be easily extended to this setting. Details on this are thus omitted.
\section{Application to antihyperglycemic drugs on weight gain}
We now apply our proposed methods to investigate the risk of moderate to severe weight gain of metformin versus sulfonylureas as initial therapy for new users of antihyperglycemic drugs (i.e. prescribed for patients with diabetes) during the period of 1995 to 2011. The data for this analysis were extracted from The Health Improvement Network 
\citep{lewis2007validation}. From this database, we select patients who were present for at least 180 days before receiving any antihyperglycemic drugs, and then were started on an initial therapy with either metformin $(D=1)$ or a sulfonylurea $(D=0)$, with a baseline glycosylated hemoglobin (HbA1c) of $\ge 7$\% \citep{ertefaie2017tutorial}. 
The outcome of interest $Y$ is a binary variable, which indicates an increase of at least 10\% of BMI at two years of follow-up compared to each patient's baseline. This cut-off value is chosen based on the definition of moderate-to-severe weight gain (i.e. Grade 2-3) of the common terminology criteria for adverse events \citep{savarese2013common}. Although the cut-offs are proposed for weight, we use the same thresholds for BMI as this measure is a linear function of weight. The frequency of the outcome among patients treated with metformin and sulfonylurea is 3.6\% and 11.7\%, respectively.

During the research period, the use of metformin rose very quickly, while the use of sulfonylureas declined quite dramatically. Beginning in 2000, metformin became more commonly used than sulfonylurea. We thus choose the time point $t=0$ to be the period of 1995 to 1999 (i.e. sulfonylurea more commonly used), and $t=1$ to be the period of 2000 to 2011 (i.e. metformin more commonly used). We make the assumption that at each time point, a random sample of patients was taken from the population of interest (i.e. Assumption 4).
Besides, the aforementioned variability in the prescription trends of both drugs also led us to define our instrument for difference-in-differences based on provider preference. For this, we first calculated the proportion of patients starting on metformin within each general practitioner practice in 1995.  We then assigned $Z=1$ if this proportion is larger than the median of all practices and $Z=0$ otherwise. We did not consider baseline covariate adjustment in this analysis.
\begin{table}[htb!!]
\centering
\caption{Data characteristics across two timepoints}
\scalebox{0.9}{\begin{tabular} { lcc}
\textbf{Characteristics} & \textbf{Time 0}& \textbf{Time 1}\\
Number of patients & 1656 & 15234 \\
$P(D_t=1)$ & 0.46 & 0.86 \\
$P(Z=1|T=t)$ & 0.58 & 0.53 \\
$P(Y_t = 1|D_t = 1)$ & 0.03 & 0.04\\
$P(Y_t = 1|D_t = 0)$ & 0.10 & 0.12 
\end{tabular}}
\label{tab1}
\end{table}

Data from 16890 patients (117 practices) are finally included. By solving the sample analog of equation (\ref{eq12}), we obtain an estimate of of $\beta = -1.27$ for the treatment effect on the log relative risk scale, with a 95\% confidence interval ranging from $ -3.07$ to $0.53$. This suggests that the risk of moderate to serious weight gain from metformin is $e^{-1.27}=0.281$ times as low compared with sulfonylurea. Although this finding is not statistically significant, the direction of the result agrees with prior findings, which also suggests an increase risk of weight gain by sulfonylurea compared to other oral antihyperglycemic drugs \citep{phung2010effect,confederat2016side}. Here we focus though on the incidence of  moderate to severe weight gain. 
\section{Conclusion}
In this paper, we have proposed novel additive and multiplicative structural mean models for the instrumented difference-in-differences design. By applying semi-parametric theory, we also develop multiple estimation approaches for the parameters indexing such models, thereby enabling the estimation of the average exposure effect in the whole population or among the exposed, on the additive and multiplicative scales. The suggested methods can be used in continuous outcome settings (additive structural mean models), or count outcome settings (multiplicative structural mean models). 
In the special case where the outcome indicates a rare event with a small success probability (i.e. around $10\%$ or less as a rule of thumb), the multiplicative structural mean models can also be good approximations for the true structural mean models that have a logistic link function. However, the estimation of the treatment effect in this setting often requires a quite large sample size to obtain valid inference.

A potential direction for future research is to develop estimation strategies for structural mean models with a logistic link function, without having to assume the binary outcome is rare. The difficulty with constructing consistent estimators for such logistic models is in finding a residual $\epsilon( O, \beta)$ satisfying a moment condition similar to (\ref{moment.rr}), i.e. $E\{\epsilon( O, \beta)\mid X, Z\} = 0$. Extension to a logistic link may thus require a rather different line of thinking. Finally, while we here focus on two time points, the proposed models should also be extended to multiple time points settings where many additional complications may also present, such as staggered treatment adoption, violation of parallel trend assumptions and so forth \citep{roth2022s}. 

\bibliographystyle{biorefs}   
\bibliography{main}

\section*{Supplementary Materials}
\appendix
\section{Identification results}
\subsection{Additive SMMs for IDiD}
We first consider the following additive SMM for IDiD to estimate the average treatment effect $\beta(\bm x)$:
\begin{align*}
E(Y_1^{d^*} - Y_0^{d}|\bm X = \bm x, Z) =  \beta(\bm x) \cdot (d^* - d) + m(\bm x), 
\end{align*}
for all $d,d^*$. 
From this model we have that $\forall d, d^*$:
\begin{align*}
    E(Y_1^{d^*}  - \beta(\bm X)d^*|\bm X, Z) &= 
    E(Y_0^{d} - \beta(\bm X)d|\bm X, Z) + m(\bm X)\\
    E\{E(Y_1^{d^*}  - \beta(\bm X)d^*|U_0, U_1, D_0,\bm X, Z)|\bm X, Z\} &= 
    E\{E(Y_0^{d} - \beta(\bm X)d|U_0, \bm X, Z) | \bm X, Z\}+ m(\bm X)\\
    E\{E(Y_1^{d^*}  - \beta(\bm X)d^*|U_0, U_1, D_0,D_1=d^*,\bm X, Z)|\bm X, Z\} &= 
    E\{E(Y_0^{d} - \beta(\bm X)d|U_0, D_0 = d, \bm X, Z) | \bm X, Z\}+ m(\bm X)\\
    E\{E(Y_1  - \beta(\bm X)D_1|U_0, U_1, D_0,D_1=d^*,\bm X, Z)|\bm X, Z\} &= 
    E\{E(Y_0 - \beta(\bm X)D_0|U_0, D_0 = d, \bm X, Z) | \bm X, Z\}+ m(\bm X)
\end{align*}
The third equality follows from assumption (2i), i.e. sequential ignorability. The last one follows from assumption (1), i.e. consistency. As this holds for all values of $d$ and $d^*$, one can rewrite it as:
\begin{align*}
    E\{E(Y_1  - \beta(\bm X)D_1|U_0, U_1, D_0,D_1,\bm X, Z)|\bm X, Z\} &= 
    E\{E(Y_0 - \beta(\bm X)D_0|U_0, D_0, \bm X, Z) | \bm X, Z\}+ m(\bm X)\\
    E(Y_1  - \beta(\bm X)D_1|\bm X, Z) &= 
    E(Y_0 - \beta(\bm X)D_0 | \bm X, Z)+ m(\bm X)\\
    E(Y_1-Y_0|\bm X, Z) &= \beta(\bm X)\cdot E(D_1 - D_0|\bm X, Z) + m(\bm X)
\end{align*}
By plugging $Z=1$ and $Z=0$ into the above expression, we obtain the identification result of interest.

Consider now the treatment effect among the treated $\beta'(\bm X)$. The structural mean model for $\beta'(\bm X)$ can be expressed as:
\begin{align*}
E(Y_t^d|D_t = d, \bm X = \bm x, Z) - E(Y_t^0|D_t = d,\bm X = \bm x, Z) = \beta'(\bm x) d~~~~~\mathrm{for}~~~~d=0,1
\end{align*}
One then has:
\begin{align*}
    E(Y_t^d - \beta'(\bm X)d|D_t = d, \bm X, Z) &= E(Y_t^0|D_t = d,\bm X, Z)\\
    E(Y_t - \beta'(\bm X)D_t|D_t = d, \bm X, Z) &= E(Y_t^0|D_t = d,\bm X, Z)
\end{align*}
Plugging $t=1$ and $t=0$ into the above expression, one then has:
\[E(Y_1 - Y_0|\bm X, Z) - \beta'(\bm X) E(D_1-D_0|\bm X, Z) = E(Y_1^0 - Y_0^0|\bm X, Z) = E(Y_1^0 - Y_0^0|\bm X), \]
where the second equality follows from assumption 3 in section 2. Plugging $Z=1$ and $Z=0$ into this equation, we obtain the identification result of interest.

\subsection{Multiplicative SMMs for IDiD}
The proof for the multiplicative SMM for IDiD follows the same steps. For all $d$ and $d^*$:
\begin{align*}
E(Y_1^{d^*}|\bm X, Z) &= E(Y_0^{d}|\bm X, Z)e^{\beta(\bm X) \cdot (d^* - d) + m(\bm X)}\\
E(Y_1^{d^*}e^{-\beta(\bm X)d^*}|\bm X, Z) &= E(Y_0^{d}e^{-\beta(\bm X)d + m(\bm X)}|\bm X, Z)\\
E\{E(Y_1^{d^*}e^{-\beta(\bm X)d^*}|U_0,U_1,D_0,\bm X, Z) |\bm X, Z\}&= E(Y_0^{d}e^{-\beta(\bm X)d + m(\bm X)}|U_0,\bm X, Z)|\bm X, Z\}\\
E\{E(Y_1^{d^*}e^{-\beta(\bm X)d^*}|U_0,U_1,D_0,D_1=d^*,\bm X, Z) |\bm X, Z\}&= E(Y_0^{d}e^{-\beta(\bm X)d + m(\bm X)}|U_0,D_0=d,\bm X, Z)|\bm X, Z\}\\
E\{E(Y_1e^{-\beta(\bm X)D_1}|U_0,U_1,D_0,D_1=d^*,\bm X, Z) |\bm X, Z\}&= E(Y_0e^{-\beta(\bm X)D_0 + m(\bm X)}|U_0,D_0=d,\bm X, Z)|\bm X, Z\}
\end{align*}
Therefore,
\begin{align*}
    E\{E(Y_1e^{-\beta(\bm X)D_1}|U_0,U_1,D_0,D_1,\bm X, Z) |\bm X, Z\}&= E(Y_0e^{-\beta(\bm X)D_0 + m(\bm X)}|U_0,D_0,\bm X, Z)|\bm X, Z\}\\
    E(Y_1e^{-\beta(\bm X)D_1}|\bm X, Z)&= E(Y_0e^{-\beta(\bm X)D_0 + m(\bm X)}|\bm X, Z)
\end{align*}
This finishes the proof.

Consider now the treatment effect among the treated $\beta'(\bm X)$ on the multiplicative scale, i.e.:
\[\beta'(\bm x) = \frac{E(Y_t^{1}|D_t = 1, \bm X = \bm x)}{E(Y_t^{0}|D_t = 1, \bm X = \bm x)}  \]
The structural mean model for $\beta'(\bm X)$ can be expressed as:
\begin{align*}
E(Y_t^d|D_t = d, \bm X, Z) = E(Y_t^0|D_t = d,\bm X, Z) e^{\beta'(\bm X) d}~~~~~\mathrm{for}~~~~d=0,1
\end{align*}
One then has:
\begin{align*}
    E(Y_t^de^{-\beta'(\bm X) d}|D_t = d, \bm X, Z) &= E(Y_t^0|D_t = d,\bm X, Z)\\
    E(Y_t e^{-\beta'(\bm X) D_t}|D_t = d, \bm X, Z) &= E(Y_t^0|D_t = d,\bm X, Z)\\
    E(Y_t e^{-\beta'(\bm X) D_t}|D_t, \bm X, Z) &= E(Y_t^0|D_t,\bm X, Z)\\
    E(Y_t e^{-\beta'(\bm X) D_t}| \bm X, Z) &= E(Y_t^0|\bm X, Z)\\
    E(Y_t e^{-\beta'(\bm X) D_t}| \bm X, Z) &= E(Y_t^0|\bm X)
\end{align*}
Plugging $T=1$ and $T=0$ into this equation, we obtain that $
    E(Y_1 e^{-\beta'(\bm X) D_1}| \bm X, Z) = E(Y_1^0|\bm X)$ and that $
    E(Y_0 e^{-\beta'(\bm X) D_0}| \bm X, Z) = E(Y_0^0|\bm X)$. As a result,
\[\frac{E(Y_1 e^{-\beta'(\bm X) D_1}| \bm X, Z)}{E(Y_0 e^{-\beta'(\bm X) D_0}| \bm X, Z)} = \frac{E(Y_1^0|\bm X)}{E(Y_0^0|\bm X)} \]
Denoting $m(\bm X) = \frac{E(Y_1^0|\bm X)}{E(Y_0^0|\bm X)}$, we obtain the identification result of interest.

When $\bm X = \empty$ and $m(\bm X) = 0$, the above identification result implies that:
\begin{align*}
    E(Y_1e^{-\beta}D_1|Z=0) &= E(Y_0e^{-\beta}D_0|Z=0)\\
    E(Y_1e^{-\beta}D_1|Z=0) &= E(Y_0e^{-\beta}D_0|Z=0)
\end{align*}
This motivates the estimation approach discussed in section 3.2.
\section{Estimation of multiplicative SMMs for IDiD}
\subsection{$m(\bm X)$ pre-specified}
The density of a single observation can be expressed as $p(\bm O) = p_0(\bm X, Z) p_1(\bm O)$, where $p_0(\bm x, z)$ is any non-negative function such that:
$\int p_0(\bm x, z)d\bm x dz = 1$
and $p_1(\bm o)$ is any non-negative function such that:
\begin{align*}
    \int p_1(\bm o)d\bm x dz = 1; \quad \quad \quad
    \int \epsilon(\bm o) p_1(\bm o) d\bm x dz =0
\end{align*}
To develop the semiparametric theory and define the semiparametric nuisance tangent space, we first consider parametric submodels. Instead of arbitrary functions $p_1(\bm O)$ and $p_0(\bm X, Z)$ satisfying the above constraints, we will consider parametric submodels $p_1(\bm O, \bm \nu_1)$ and $p_0(\bm X, Z, \bm \nu_2)$, where $\bm \nu_1$ is an $r_1$-dimensional vector and $\bm \nu_2$ is an $r_2$-dimensional vector. Thus $\bm \nu = \begin{pmatrix} \bm \nu_1^T & \bm \nu_2^T\end{pmatrix}^T$ is an $r$-dimensional vector, $r=r_1+r_2$. This parametric submodel is given as $\mathcal{P}_{\theta,\nu} = p_1(\bm O,\bm \nu_1)p_0(\bm X, Z, \bm \nu_2)$ for $\begin{pmatrix} \bm  \theta^T & \bm \nu^T\end{pmatrix}^T \in \Omega_{\bm \theta,\bm \nu} \subset \mathcal{R}^{q+r} $. Also, to be a parametric submodel, $\mathcal{P}_{\bm \theta,\bm \nu}$ must contain the truth, i.e. $p^*(\bm O) =  p_1(\bm O,\bm \nu_{10})p_0(\bm X, Z, \bm \nu_{20})$. The parametric submodel nuisance score vector is given as:
\begin{align*}
    S_\nu(\bm O, \bm \theta_0, \bm \nu_0) &= \bigg\{\bigg(\frac{\partial \log p_1(\bm O,\bm \theta,\bm \nu_1)}{\partial \bm \nu_1} \bigg)^T, \bigg(\frac{\partial \log p_0(\bm X, Z,\bm \nu_2)}{\partial\bm \nu_2} \bigg)^T
    \bigg\}^T\bigg |_{\bm \theta = \bm \theta_0, \bm \nu = \bm \nu_0}\\
    &=\{\bm S_{\bm \nu_1}^T(\bm O,\bm \theta_0,\bm \nu_{10}), \bm S_{\bm \nu_0}^T(\bm X,Z,\bm \theta_0,\bm \nu_{20}\}^T
\end{align*}
A typical element in the parametric submodel nuisance tangent space is
given by:
\[\bm B^{q\times r}\bm S_\nu(\bm O) =  \bm B_1^{q\times r_1}\bm S_{\nu_1}(\bm O) + 
\bm B^{q\times r_2}\bm S_{\nu_2}(\bm X, Z)\]
where $\bm B$ are matrices of constants.
Therefore, the parametric submodel nuisance tangent space $\Lambda_{\bm \nu} = \{\bm B^{q\times r}S_{\bm \nu}(\bm O) \}$ can be written as the direct sum of the two spaces $\Lambda_{\bm \nu_1} = \{\bm B_1^{q\times r_1}\bm S_{\nu_1}(\bm O)\}$ and $\Lambda_{\nu_2} = \{\bm B^{q\times r_2}\bm S_{\nu_2}(\bm X, Z) \}$.

The semiparametric nuisance tangent space $\Lambda$ is the mean-square closure of $\Lambda_{\bm \nu_1} \oplus \Lambda_{\bm \nu_1}$. Because $\bm \nu_1$ and $\bm \nu_2$ are variationally independent - that is, proper densities in the
parametric submodel can be defined by considering any combination of $\bm \nu_1$
and $\bm \nu_2$, this implies that $\Lambda$ is the direct sum of $\Lambda_{1s}$ and $\Lambda_{2s}$, which are the mean-square closures of all $\Lambda_{\bm \nu_1}$ and of all $\Lambda_{\bm \nu_2}$, respectively.

By theorem 4.6 in \citet{tsiatis2006semiparametric}, the space $\Lambda_{2s}$ consists of all $q$-dimensional mean-zero functions of $\bm X$ and $Z$ with finite variance. In contrast, the space $\Lambda_{1s}$ consists of all $q$-dimensional random functions $a(\bm O)$ that satisfy $E\{\bm a(\bm O)|\bm X\} = \bm 0$ and $E\{\bm a(\bm O)\epsilon|\bm X \} = \bm 0$. The proof of this is similar to that of theorem 4.7 in \citet{tsiatis2006semiparametric}.

The consequence of the above results is that any element of the nuisance tangent space $\Lambda_1$ can be written as $\bm S_0(Z,\bm X) + \bm S_1(\bm O)$, where $\bm S_0(Z,\bm X)$ and $\bm S_1(\bm O)$ are $q$-dimensional functions of $(\bm X, Z)$ and of $\bm O$ such that:
\begin{align*}
    E\{\bm S_0(Z,\bm X)\} &= \bm 0\\
    E\{\bm S_1(\bm O)|Z,\bm X\} &= \bm 0\\
    E(\epsilon(\bm O)\bm S_1(\bm O)|Z,X) &= \bm 0
\end{align*}
To find the orthocomplement of the nuisance tangent space, we take an arbitrary $q$-dimensional function $\bm h(\bm O)$ of the observed data and do an orthogonal projection. Denote $\bm S_0^p(Z,\bm X) + \bm S_1^p(Z, \bm X)$ the projection of $\bm h(\bm O)$ on the nuisance tangent space. One then has:
\[E\big[\{\bm h(\bm O) - \bm S_0^p(Z,\bm X) - \bm S_1^p(\bm O)\}^T\{\bm S_0(Z,\bm X) + \bm S_1(\bm O)\}\big] = 0 \]
for all $\bm S_0(Z,\bm X)$ and $\bm S_1(\bm O)$ obeying the restriction of the nuisance scores.

Setting $\bm S_1 = \bm 0$, we find that $E\big[\{\bm h(\bm O) - \bm S_0^p(Z,\bm X) - \bm S_1^p(\bm O)\}^T \bm S_0(Z,\bm X)\big] = 0$. This implies that:
\[E\{\bm h(\bm O) - \bm S_0^p(Z,\bm X) - \bm S_1^p(\bm O)|Z,\bm X\} = \bm 0 \]
from which $S_0^p(Z,\bm X) = E\{ h(\bm O)|Z,\bm X\}$.

Setting $\bm S_0 = \bm 0$, we find that:
\[0 =  E\big[\{\bm h(\bm O) - E\{ \bm h(\bm O)|Z,\bm X\} - \bm S_1^p(\bm O)\}^T\bm S_1(\bm O)\big]\]
for all mean-zero $q$-dimensional functions $\bm S_1(\bm O)$ given $Z,\bm X$ such that $E\{\epsilon(\bm O)\bm S_1(\bm O) |Z,\bm X\} = \bm 0$.
We thus conjecture that:
\[\bm S_1^p(\bm O) = \bm h(\bm O) - E(\bm h(\bm O)|\bm X, Z) - \epsilon \bm g(\bm X, Z)\]
where $\bm g(\bm X, Z)$ is an unknown $q$-dimensional function of $(\bm X, Z)$. For this to be a valid conjecture, we need to find $\bm g(\bm X, Z)$ such that $\bm S_1^p$ satisfies the restrictions of being a valid nuisance score, i.e. $E(\bm S_1^p(\bm O)|\bm X, Z) = \bm 0$ and $E(\epsilon(\bm O)\bm S_1^p(\bm O)|\bm X, Z) = \bm 0$. The first requirement is obviously satisfied for any function $\bm g(\bm X, Z)$. The second one is equivalent to that:
\begin{align*}
    E(\epsilon \bm h(\bm O)|\bm X, Z) = E(\epsilon^2|\bm X, Z)\bm g(\bm X, Z)
\end{align*}
This hence implies that $\bm g(\bm X, Z) = E(\epsilon \bm h(\bm O)|\bm X, Z)\cdot E^{-1}(\epsilon^2|\bm X, Z)$.

The elements of the orthogonal nuisance tangent space $\Lambda^\perp_1$ thus have the form: 
\[\bm h(\bm O) - \bm S_0^p - \bm S_1^p = \epsilon \cdot E[\epsilon \bm h(\bm O)|\bm X, Z]\cdot E^{-1}(\epsilon^2|\bm X, Z)\]
where $\bm h(\bm O)$ is an arbitrary mean-zero $q$-dimensional function of the data. We thus have $\Lambda_1^\perp = \bigg\{\bm d(\bm X, Z) \epsilon \bigg\}$ where $\bm d(\bm X, Z)$ is an arbitrary $q$-dimensional function of $(\bm X, Z)$.

Denote $\bm \phi^o = \bm d^o(\bm X, Z)\epsilon$ the efficient influence function in $\Lambda_1^\perp$, one then has $E\{\bm\phi\bm\phi^{o,T}\} = -E\bigg\{\frac{\partial \bm \phi}{\partial\bm\theta} \bigg\}$ for all $\bm\phi\in \Lambda_1^\perp$ \citep{tsiatis2006semiparametric}. As a result,
\[E\big\{\bm d(\bm X, Z)\epsilon \bm d^o(\bm X, Z)^T\epsilon \big\} = - E\bigg\{\frac{\partial (\bm d(\bm X, Z)\epsilon)}{\partial\bm\theta} \bigg\}\]
for all arbitrary $q$-dimensional functions $\bm d(\bm X, Z)$. Some algebraic transformations then give:
\[E \{\bm d(\bm X,Z) [\epsilon^2\bm d^o(\bm X, Z)^T + (\epsilon'_{\bm\theta})^T]\} = \bm 0\]
As this needs to hold for all $\bm d(\bm X,Z)$, one then has $E[\epsilon^2\bm d^o(\bm X, Z)^T + (\epsilon'_{\bm\theta})^T|\bm X, Z] = \bm 0$. This implies that:
\[\bm d^o(\bm X, Z) = -E(\epsilon'_{\bm\theta}|\bm X, Z)\cdot E^{-1}(\epsilon^2|\bm X, Z)\]
The efficient influence function is thus $\bm\phi^o = -E(\epsilon'_{\bm\theta}|\bm X, Z)\cdot E^{-1}(\epsilon^2|\bm X, Z) \cdot \epsilon$.
\subsection{$m(\bm X)$ unspecified}
\subsubsection{Derivation of the orthogonal nuisance tangent space}
Assume first that $m(\bm X)$ is known. By a similar proof as in the previous section, one can show that the orthogonal complement of the nuisance tangent space is $\Lambda_1^{\perp} = \{\bm d(\bm X, Z)\epsilon\}$, where $\bm d(\bm X, Z)$ satisfies $ E\bigg\{\bm d^{k\times1}(\bm X, Z) \bigg(\frac{\partial \epsilon}{\partial \bm\theta}\bigg)^T\bigg\} = I^{k\times k}$. 

To regconize that $m(\bm X)$ is unknown, we now take the subspace of $\Lambda_1^\perp$ that is orthogonal to the nuisance scores for $m(\bm X)$. For this, consider a parametric submodel $m(\bm X, \bm t)$ where $\bm t$ is an $r_3$-dimensional vector. The score $S_t(\bm O) = \frac{df(D_0, Y_0, D_1, Y_1|\bm X, Z, t)}{dt}\big|_{t=t_0}$ of this parametric submodel then satisfies that:
\[E\big(\epsilon \bm S_{\bm t}(\bm O)|Z, \bm X\big)=E\big(\epsilon'_{\bm t = \bm t_0} | Z, \bm X\big) = e^{m(\bm X)}\frac{dm_{\bm t}(\bm X)}{dt}\bigg|_{\bm t= \bm t_0}E(Y_0e^{-\beta(\bm X) D_0}|Z, \bm X)\]
where $\epsilon'_{\bm t= \bm t_0} = \frac{d\epsilon}{dt}\big|_{\bm t= \bm t_0}$
Hence, $S_t(\bm O)$ must satisfy:
\begin{align*}
    E(\epsilon S_t(\bm O)|Z,\bm X) = E(\epsilon S_t(\bm O)|\bm X) \frac{E(Y_0e^{-\beta(\bm X) D_0}|Z,\bm X)}{E(Y_0e^{-\beta(\bm X) D_0}|\bm X)}
\end{align*}
A typical element in the parametric submodel nuisance tangent space is given by $B^{k\times r_3}\bm S_t(\bm O)$. Thus, the semiparametric nuisance tangent space $\Lambda_2$ is the mean-square closure of $\Lambda_t = \{ B^{k\times r_3}\bm S_t(\bm O)\}$. It can be proven that $\Lambda_2$ consists of all $k$-dimensional mean-zero functions $\bm b(\bm O)$ that satisfy:
\begin{align*}
    E(\epsilon \bm b(\bm O)|Z,\bm X) = E(\epsilon \bm b(\bm O)|\bm X) \frac{E(Y_0e^{-\beta(\bm X) D_0}|Z,\bm X)}{E(Y_0e^{-\beta(\bm X) D_0}|\bm X)}
\end{align*}

We are now ready to project $\bm d(\bm X, Z)\epsilon$ orthogonal to $\Lambda_2$. In particular, we need:
\[E\big\{[\bm d(Z,\bm X)\epsilon - \bm b^p(\bm O)]^T\bm b(\bm O)\big\} = 0\]
for all $\bm b(\bm O) \in \Lambda_2$. Here, $\bm b^p(\bm O)$ is the projection of $\bm d(Z,\bm X)\epsilon$ on $\Lambda_2$. As $\bm b ^p(\bm O)$ is an element of $\Lambda_1^\perp$ found in the previous section, $\bm b^p(\bm O) = \bm d^*(Z,\bm X)\epsilon$ for some function $\bm d^*(\bm X, Z)$. This implies that:
\begin{align*}
    0 &= E\big\{\big[d(Z,\bm X) - d^*(Z,\bm X)\big]^T\epsilon \bm b(\bm O) \big\}\\
    &=  E\big\{ 
    E\big[\big(d(Z,\bm X) - d^*(Z,\bm X)\big)\lambda(\bm X, Z)\big|\bm X\big]^T\epsilon \bm (\bm O) \big\}
\end{align*}
for all $\bm b(\bm O)$ and $\lambda(\bm X, Z)=\frac{E(Y_0e^{-\beta(\bm X) D_0}|Z,\bm X)}{E(Y_0e^{-\beta(\bm X) D_0}|\bm X)}$. Hence, $\bm d^*(Z,\bm X)$ must satisfy that:
\begin{align}\label{c1}
    E\big[\big(\bm d(Z,\bm X) - \bm d^*(Z,\bm X)\big)\lambda(\bm X, Z)\big|\bm X\big] = \bm  0
\end{align}
Besides, for this to be a valid projection, we also need that $\bm b^p(\bm O)$ obeys the score function restriction. In other words,
\[E\big[\epsilon^2 \bm d^*(Z,\bm X) \big | Z, \bm X \big] = \lambda(Z,\bm X) E\big[\epsilon^2 \bm d^*(Z,\bm X) \big| \bm X\big]\]
or with $\sigma^2(Z,\bm X) = E(\epsilon^2|Z, \bm X)$:
\begin{align}\label{c2}
  \sigma^2(Z,\bm X) \bm d^*(\bm X, Z) = \lambda(\bm X, Z) E\{\sigma^2(Z,\bm X) \bm d^*(\bm X, Z) |\bm X\}   
\end{align}
It can be shown that the function $\bm d^*(\bm X, Z)$ satisfying the conditions (\ref{c1}) and (\ref{c2}) is:
\[\bm d^*(\bm X, Z) = \frac{\lambda(\bm X, Z)\sigma^{-2}(\bm X, Z)}{E(\lambda(\bm X, Z)\sigma^{-2}(\bm X, Z)|\bm X)} E\big(\bm d(\bm X, Z)\lambda(\bm X, Z)|\bm X\big ) \]
We thus conclude that the orthogonal nuisance tangent space of the proposed SMMs with $m(\bm X)$ unknown is:
\[\Lambda^{\perp}_2 = \bigg\{\bigg[\bm d(\bm X, Z) -  \frac{\lambda(\bm X, Z)\sigma^{-2}(\bm X, Z)}{E(\lambda(\bm X, Z)\sigma^{-2}(\bm X, Z)|\bm X)} E(\bm d(\bm X, Z)\lambda(\bm X, Z)|\bm X) \bigg]\epsilon\bigg\} \]

To verify our derivation, we double check that the estimating functions in $\Lambda_2^{\perp}$ have mean zero even when $m(\bm X)$ is mis-specified, i.e. $m(\bm X) \ne m_0(\bm X)$, where $m_0(\bm X)$ denotes the true form of $m(\bm X)$ that is unknown. 

Because $E\{\epsilon(m_0)|Z,\bm X\} = 0$, \[E(Y_1 e^{-\beta(\bm X) D_1}|\bm X, Z) = E(Y_0 e^{-\beta(\bm X) D_0}|\bm X, Z)m_0(\bm X)\] 
Denote $\bm f(m) = \big[\bm d(\bm X, Z) - \bm d^*(\bm X, Z) \big] \epsilon (m)$ an arbitrary element of $\Lambda_2^\perp$, we have:
\[ E\{\bm f(m)\} = E \bigg\{\big[ \bm d(\bm X, Z) - \bm d^*(\bm X, Z) \big] \cdot E\big[\epsilon(m)|\bm X, Z\big] \bigg\} \]
where:
\begin{align*}
E\big[\epsilon(m)|\bm X, Z\big] &= E\big[Y_1e^{-\beta(\bm X) D_1}|\bm X,Z\big] - E\big[Y_0e^{-\beta(\bm X) D_0 + m(\bm X)}|\bm X, Z\big]\\[5pt]
& = E\big[Y_0e^{-\beta(\bm X) D_0}|\bm X, Z\big]\cdot\big[ m_0(\bm X) - m(\bm X)\big]\\[5pt]
& = \lambda(\bm X, Z) \cdot E\big[Y_0e^{-\beta(\bm X) D_0}|\bm X \big] \cdot \big[ m_0(\bm X) - m(\bm X)\big].
\end{align*}
Denote $\omega(\bm X) = E\big[Y_0e^{-\beta(\bm X) D_0}|\bm X \big] \cdot \big[ m_0(\bm X) - m(\bm X)\big]$, we then have:
\begin{align*}
 E\{\bm f(m)\} &=  E\bigg\{ \bigg[ \bm d(\bm X, Z) - \frac{\lambda(\bm X, Z)\sigma^{-2}(\bm X, Z)}{E\big[\lambda^2(\bm X, Z)\sigma^{-2}(\bm X, Z)|\bm X\big]}\cdot E(\bm d(\bm X, Z)\lambda(\bm X, Z)|\bm X)\bigg]\cdot \lambda(\bm X, Z)  \cdot \omega (\bm X) \bigg\} \\
 &= \bm 0.
\end{align*}
This explains the double robustness property of the proposed approaches.  

\subsubsection{Influence function for $\theta = e^{-\beta} - 1$ based on equation (\ref{eq.nonpara})}
Assume that $\beta(\bm X) = \beta$, which means $\bm X$ are not effect modifiers and $k=1$. Denote $\theta := e^{-\beta}-1$. We have that $\theta$ can be defined as the solution of equation (\ref{eq.nonpara}).
One then has:
\[
 \int \bigg[ \underbrace{d(\bm x, z) - \frac{\theta a_1(x,\mathcal{P}) + a_2(x,\mathcal{P})}{\theta a_3(x,\mathcal{P}) +  a_4(x,\mathcal{P})}}_\text{$A(o,\mathcal{P})$} \bigg]
 \bigg[\underbrace{(y_1d_1-y_0d_0)\theta + y_1 - y_0 }_\text{$B(o,\mathcal{P})$}\bigg]  d\mathcal{P} = 0
\]
Evaluate the left-hand side at $\mathcal{P}_t = (1-t)\mathcal{P} + t\mathbbm{1}_{\tilde{o}}$, where $\mathbbm{1}_{\tilde{o}}$ denotes the Dirac delta function at $\tilde{o}$, one then has:
\begin{align} \label{eq18}
 E_\mathcal{P}\bigg[ \bigg(\frac{d}{dt} A(O)\bigg|_{t=0}\bigg) B(O) + A(O)\bigg(\frac{d}{dt} B(O)\bigg|_{t=0}\bigg)\bigg] + \int A(o,\mathcal{P})B(o,\mathcal{P})\frac{d\mathcal{P}_t}{dt}\bigg|_{t=0} = 0
 \end{align}
Besides,
\begin{align*}
\frac{d}{dt} A(o)\bigg|_{t=0} &= -\frac{\theta'_{t=0}a_1 + a'_{1,t=0}\theta + a'_{2,t=0}}
{\theta a_3 + a_4} + \frac{\theta a_1 + a_2}{(\theta a_3 + a_4)^2}(\theta'_{t=0}a_3 + a'_{3,t=0}\theta + a'_{4,t=0})\\
&= \frac{a_2a_3-a_1a_4}{(\theta a_3 + a_4)^2}(\theta'_{t=0}) - \frac{a'_{1,t=0}\theta + a'_{2,t=0}}{\theta a_{3} + a_{4}} + \frac{(\theta a_1 + a_2)(a'_{3,t=0}\theta + a'_{4,t=0})}{(\theta a_3 + a_4)^2}
\end{align*}
where $\theta'_{t=0} = \frac{d}{dt}\theta(\mathcal{P}_t)\bigg |_{t=0}$ and $a'_{i,t=0} = \frac{d}{dt}a_i(x,\mathcal{P}_t)\bigg |_{t=0}$ for $i = \overline{1,4}$. Also, $a_i$ is short-hand notation for $a_i(x,\mathcal{P})$ for $i = \overline{1,4}$. We also have:
\begin{align*}
\frac{d}{dt} B(o)\bigg|_{t=0} &= \theta'_{t=0}(y_1d_1 - y_0d_0)
\end{align*}
Finally,
\begin{align*}
\int  A(o,\mathcal{P})B(o,\mathcal{P})\frac{d\mathcal{P}_t}{dt}\bigg|_{t=0}  &= \int A(o,\mathcal{P}) B(o,\mathcal{P})\big[ \mathbbm{1}_{\tilde{o}} - f(o)\big] do\\
&= A(\tilde{o},\mathcal{P}) B(\tilde{o},\mathcal{P}) - \int A(o)B(o)f(o)do\\
& = A(\tilde{o},\mathcal{P}) B(\tilde{o},\mathcal{P}) 
\end{align*}
Plugging in the above results in (\ref{eq18}), we then have:
\begin{align*} 
 0 &= E_\mathcal{P}\bigg[\bigg( \frac{a_2a_3-a_1a_4}{(\theta a_3 + a_4)^2}(\theta'_{t=0}) - \frac{a'_{1,t=0}\theta + a'_{2,t=0}}{\theta a_{3} + a_{4}} + \frac{(\theta a_1 + a_2)(a'_{3,t=0}\theta + a'_{4,t=0})}{(\theta a_3 + a_4)^2}\bigg) B(O) \bigg]\\
 &+ E_\mathcal{P}\bigg[A(O)(\theta'_{t=0})(Y_1D_1 - Y_0D_0)\bigg] +  A(\tilde{o},\mathcal{P}) B(\tilde{o},\mathcal{P})
 \end{align*}
 Taking $\theta'_{t=0}$ out of the expectation, one then has:
 \begin{align*} 
& (\theta'_{t=0}) E_\mathcal{P}\bigg[\bigg( \frac{a_2a_3-a_1a_4}{(\theta a_3 + a_4)^2}B(O) + A(O)(Y_1D_1-Y_0D_0) \bigg)\bigg] \\
& = E_\mathcal{P}\bigg[\bigg(\frac{a'_{1,t=0}\theta + a'_{2,t=0}}{\theta a_{3} + a_{4}} - \frac{(\theta a_1 + a_2)(a'_{3,t=0}\theta + a'_{4,t=0})}{(\theta a_3 + a_4)^2}\bigg) B(O) \bigg] -  A(\tilde{o},\mathcal{P}) B(\tilde{o},\mathcal{P})
 \end{align*}
 Hence,
  \begin{align*} 
 \theta'_{t=0} = \frac{E_\mathcal{P}\bigg[\bigg(\frac{a'_{1,t=0}\theta + a'_{2,t=0}}{\theta a_{3} + a_{4}} - \frac{(\theta a_1 + a_2)(a'_{3,t=0}\theta + a'_{4,t=0})}{(\theta a_3 + a_4)^2}\bigg) B(O) \bigg] -  A(\tilde{o},\mathcal{P}) B(\tilde{o},\mathcal{P})}
{E_\mathcal{P}\bigg[\bigg( \frac{a_2a_3-a_1a_4}{(\theta a_3 + a_4)^2}B(O) + A(O)(Y_1D_1-Y_0D_0) \bigg)\bigg]}
 \end{align*}
We also have:
\begin{align*}
E_\mathcal{P}\bigg[\frac{a'_{1,t=0}\theta}{\theta a_{3} + a_{4}} B(O) \bigg] 
&= \theta \int \frac{B(\overline{y}_1,\overline{d}_1)}{\theta a_3(x) + a_4(x)}\cdot \frac{ \mathbbm{1}_{\tilde{x}}(x)}{f(x)}\big\{\tilde{y}_0\tilde{d}_0 d({x},\tilde{z}) - a_1(x)\big\}\cdot f(y_1,y_0,d_1,d_0,x)d\overline{y}_1d\overline{d}_1dx\\
&= \theta  \int B(\overline{y}_1,\overline{d}_1) \cdot
\frac{\tilde{y}_0\tilde{d}_0 d(\tilde{x},\tilde{z}) - a_1(\tilde{x})}{\theta a_3(\tilde{x}) + a_4(\tilde{x})} 
\cdot f(y_1,y_0,d_1,d_0|X=\tilde{x})d\overline{y}_1d\overline{d}_1\\
&= \frac{\theta\big[\tilde{y}_0\tilde{d}_0 d(\tilde{x},\tilde{z}) - a_1(\tilde{x})\big]}{\theta a_3(\tilde{x}) + a_4(\tilde{x})} 
 \cdot E\{ B(O)|X=\tilde{x}\}
\end{align*}\
where the first equality results from the fact that:
\[ a'_{1,t=0}(x) = \frac{ \mathbbm{1}_{\tilde{x}}(x)}{f(x)}\big\{\tilde{y}_0\tilde{d}_0 d({x},\tilde{z}) - a_1(x)\big\} \]
In a similar way, one can prove that:
\begin{align*}
E_\mathcal{P}\bigg[\frac{a'_{2,t=0}}{\theta a_{3} + a_{4}} B(O) \bigg] &= 
\frac{\tilde{y}_0 d(\tilde{x},\tilde{z}) - a_2(\tilde{x})}{\theta a_3(\tilde{x}) + a_4(\tilde{x})} 
 \cdot E\{ B(O)|X=\tilde{x}\}\\
E_\mathcal{P}\bigg[ \frac{(\theta a_1 + a_2)a'_{3,t=0}\theta}{(\theta a_3 + a_4)^2}B(O)
\bigg] &= 
\frac{(\theta a_1(\tilde{x}) + a_2(\tilde{x}))\theta}{(\theta a_3(\tilde{x}) + a_4(\tilde{x}))^2} (\tilde{y}_0\tilde{d}_0 - a_3(\tilde{x})) 
\cdot E\{ B(O)|X=\tilde{x}\}\\
E_\mathcal{P}\bigg[ \frac{(\theta a_1 + a_2)a'_{4,t=0}}{(\theta a_3 + a_4)^2}B(O)
\bigg] &= 
\frac{\theta a_1(\tilde{x}) + a_2(\tilde{x})}{(\theta a_3(\tilde{x}) + a_4(\tilde{x}))^2} (\tilde{y}_0 - a_4(\tilde{x})) 
\cdot E\{ B(O)|X=\tilde{x}\}\\
\end{align*}
\subsubsection{Remainder term for $\theta := e^{-\beta}-1$ based on equation (\ref{eq.nonpara})}
In section B.2.2, we showed that the influence function of $\theta$ is:
 \begin{align*}
 &\phi(\bm O, \theta,\bm\eta) \\
 &=\frac{1}{C}\bigg[ \bigg\{ \frac{\theta Y_0D_0d(X,Z)- a_1\theta + Y_0d(X,Z)-a_2}{\theta a_3 + a_4} - \frac{(\theta a_1 + a_2)(\theta Y_0D_0 - \theta a_3 + Y_0 - a_4)}{(\theta a_3 + a_4)^2}\bigg\} E\big\{B(\bm O)|X \big\} \\
 &- A(\bm O) B(\bm O)\bigg] 
 \end{align*}
 where $\bm\eta = c(a_1,a_2,a_3,a_4)$ is the vector of all nuisance parameters and
 \[C=C(\bm O,\theta,\eta) =  E\bigg[\bigg( \frac{a_2a_3-a_1a_4}{(\theta a_3 + a_4)^2}B(O) + A(O)(Y_1D_1-Y_0D_0) \bigg)\bigg]\]
Let $\hat{\eta} = (\hat a_1, \hat a_2, \hat a_3, \hat a_4, \hat a_5, \hat a_6)$ denote an estimate for $\eta$. Assume that $\hat{\eta}$ converges in probability to some $ \eta'$ that might be potentially different from the true value of the nuisance parameter $ \eta$. The aim is to prove that the remainder term: 
\[R(\eta,\eta') := \theta(\eta') - \theta(\eta) + E\{\phi(O,\eta')\}\]
is a second order term involving only products of the type $E[ c(\eta,\eta_1)\{f(\eta')-f(\eta)\}\{g(\eta')-g(\eta)\}]$. For this, denote $C' = C(\bm O, \theta',\eta'), A' = A(\bm O, \theta',\eta'), B' = B(\bm O, \theta', \eta')$. One then has:
\begin{align*}
&E\{ \phi(\eta')\} \\
&= \frac{1}{C'} E\bigg[ \bigg\{ \frac{Y_0D_0d(X,Z)\theta'+Y_0d(X,Z)- (a'_1\theta' + a'_2)}{\theta' a'_3 + a'_4} - \frac{(\theta a'_1 + a'_2)(\theta' Y_0D_0 - \theta' a'_3 + Y_0 - a'_4)}{(\theta' a'_3 + a'_4)^2}\bigg\} E\big\{B'|X \big\} - A' B'\bigg] \\
&=  \frac{1}{C'} E\bigg[ \bigg\{ \frac{a_1\theta' + a_2 - a'_1\theta' - a'_2}{\theta' a'_3 + a'_4} - \frac{(\theta a'_1 + a'_2)(\theta' a_3 - \theta' a'_3 + a_4 - a'_4)}{(\theta' a'_3 + a'_4)^2}\bigg\} E\big\{B'|X \big\}\bigg]  -  \frac{1}{C'} E(A' B') \\
&=  \frac{1}{C'} \underbrace{E\bigg[ \bigg\{ \frac{\textcolor{red}{(a_1\theta'- a'_1\theta' + a_2 - a'_2)(\theta'a_3'+a_4')-(\theta' a'_1 + a'_2)[\theta' (a_3 - a'_3) + a_4 - a'_4]}}{(\theta' a'_3 + a'_4)^2}\bigg\} E\big\{B'|X \big\}\bigg]}_\text{$T_1$}  -  \frac{1}{C'} \underbrace{E(A' B')}_\text{$T_2$} 
\end{align*}
where the second equality results from the law of total expectation (given $X$). The red term equals:
\begin{align*}
&\theta'(a_1 - a_1')(\theta'a_3'+a_4') 
+ \theta'a_2a_3' +a_2a_4'- a_2'\theta'a_3' - a_2'a_4' \\
&\quad - \bigg[\theta'^2 a'_1 (a_3 - a_3') + \theta'a_1'(a_4 - a_4')
+ \theta'a_2'a_3 - \theta'a_2'a_3' + a_2'a_4 - a_2'a_4'\bigg]\\
=~& \theta'(a_1 - a_1')(\theta'a_3'+a_4') 
+ \theta'(a_2a_3' - a_2'a_3) + a_2a_4' - a_2'a_4 \\
&\quad- \bigg[\theta'^2 a'_1 (a_3 - a_3') + \theta'a_1'(a_4-a_4')\bigg]\\
=~& \theta'(a_1 - a_1')(\theta'a_3'+a_4')  + \theta'a_2(a_3'-a_3) + \theta'a_3(a_2-a_2') + a_2(a_4'-a_4)+a_4(a_2-a_2')\\
&\quad- \bigg[\theta'^2 a'_1 (a_3 - a_3') + \theta'a_1'(a_4-a_4')\bigg]\\
=~ & \theta'(a_1 - a_1')(\theta'a_3'+a_4') + (\theta'a_2+\theta'^2a_1')(a_3'-a_3) + (\theta'a_3+a_4)(a_2-a_2') + (a_2 + \theta'a_1')(a_4'-a_4)
\end{align*}
Besides,
$E(B'|X) = \theta'(a_5 - a_3) + a_6 - a_4
$. 
Hence,
\begin{align*}
T_1 &=E\bigg\{ \frac{\theta'(a_1 - a_1')(\theta'a_3'+a_4') + (\theta'a_2+\theta'^2a_1')(a_3'-a_3) + (\theta'a_3+a_4)(a_2-a_2') + (a_2 + \theta'a_1')(a_4'-a_4)}{(\theta' a'_3 + a'_4)^2}\\
&\quad \times \big[\theta'(a_5 - a_3) + a_6 - a_4\big]\bigg\}\\
& = E\bigg\{\frac{U_1U_2 (\theta a_3 + a_4) }{(\theta'a_3' + a_4')^2(\theta a_3 + a_4)} \bigg\}
\end{align*}
where $U_1$ denotes the red term and $U_2 =E(B'|X)$. We now focus on $T_2$:
\begin{align*}
T_2 &= E\bigg \{\bigg[d - \frac{\theta'a_1' + a_2'}{\theta'a_3' + a_4'} \bigg]\bigg[(Y_1D_1 - Y_0D_0)\theta' + Y_1 - Y_0 \bigg] \bigg\}\\
&= \theta' E\bigg[(Y_1D_1 - Y_0D_0)d\bigg] + \textcolor{blue}{E\bigg(Y_1d - Y_0d\bigg)} 
- E\bigg\{ \frac{\theta'a_1' + a_2'}{\theta'a_3' + a_4'} \bigg[\theta'(a_5 - a_3) + a_6 - a_4\bigg]
\bigg\}
\end{align*}
Besides, from the moment condition (\ref{eq.nonpara}), we have:
\begin{align*}
0 & = E[A(\eta)B(\eta)]\\
& = E\bigg\{d\big[(Y_1D_1-Y_0D_0)\theta + Y_1 - Y_0\big]- \frac{\theta a_1 + a_2}{\theta a_3 + a_4}\big[(Y_1D_1-Y_0D_0)\theta + Y_1 - Y_0\big]\bigg\}\\
&= \theta E\bigg[ d(Y_1D_1 - Y_0D_0)\bigg] + E\bigg(Y_1d - Y_0d\bigg) - E\bigg\{\frac{\theta a_1 + a_2}{\theta a_3 + a_4} \big[(a_5-a_3)\theta + a_6 - a_4\big]\bigg\}
\end{align*}
Hence,
\[  \textcolor{blue}{E\bigg(Y_1d - Y_0d\bigg)} =  E\bigg\{\frac{\theta a_1 + a_2}{\theta a_3 + a_4} \big[(a_5-a_3)\theta + a_6 - a_4\big]\bigg\} - \theta E\bigg[ d(Y_1D_1 - Y_0D_0)\bigg] \]
Plugging this into $T_2$, one then has:
\begin{align*}
T_2 & = (\theta'-\theta) E\bigg[(Y_1D_1 - Y_0D_0)d\bigg]
- E\bigg\{ \frac{\theta'a_1' + a_2'}{\theta'a_3' + a_4'} \bigg[\theta'(a_5 - a_3) + a_6 - a_4\bigg]
\bigg\}  + E\bigg\{\frac{\theta a_1 + a_2}{\theta a_3 + a_4} \big[(a_5-a_3)\theta + a_6 - a_4\big]\bigg\}\\
&= \textcolor{purple}{(\theta'-\theta) E\bigg[(Y_1D_1 - Y_0D_0)d\bigg]} 
- \textcolor{orange}{E\bigg\{ \frac{\theta'a_1' + a_2'}{\theta'a_3' + a_4'} \theta'(a_5 - a_3)\bigg\}} 
- E\bigg\{ \frac{\theta'a_1' + a_2'}{\theta'a_3' + a_4'} (a_6 - a_4)
\bigg\}  \\
& + E\bigg\{\frac{\theta a_1 + a_2}{\theta a_3 + a_4} \big[(a_5-a_3)\theta + a_6 - a_4\big]\bigg\}
\end{align*}
Because $R(\eta, \eta') = \frac{C'(\theta'-\theta) + T_1 - T_2}{C'},$ we now focus on expanding $C'(\theta' - \theta)$:
\begin{align*}
C'(\theta' - \theta) & = (\theta'-\theta) E\bigg[\bigg( \frac{a_2'a_3'-a_1'a_4'}{(\theta a'_3 + a'_4)^2}B(\eta') + A(\eta')(Y_1D_1-Y_0D_0) \bigg)\bigg]\\
& = E\bigg\{ (\theta'-\theta)U_2 \frac{a_2'a_3'-a_1'a_4'}{(\theta a'_3 + a'_4)^2} \bigg\} 
+ E\bigg\{ (\theta'-\theta)U_2\bigg[ d(Y_1D_1-Y_0D_o) - \frac{\theta'a_1'+a_2'}{\theta'a_3'+a_4'}(Y_1D_1-Y_0D_0)
\bigg]\bigg\}\\
& = E\bigg\{ (\theta'-\theta) U_2\frac{a_2'a_3'-a_1'a_4'}{(\theta a'_3 + a'_4)^2} \bigg\} + (\theta'-\theta) E\bigg[(Y_1D_1 - Y_0D_0)d\bigg]  - (\theta' - \theta) E\bigg\{\frac{\theta'a_1'+a_2'}{\theta'a_3'+a_4'}(a_5-a_3) \bigg\}\\
&= E\bigg\{ (\theta'-\theta) U_2\frac{a_2'a_3'-a_1'a_4'}{(\theta a'_3 + a'_4)^2} \bigg\} + \textcolor{purple}{(\theta'-\theta) E\bigg[(Y_1D_1 - Y_0D_0)d\bigg]}\\
& - \textcolor{orange} {E\bigg\{\frac{\theta'a_1'+a_2'}{\theta'a_3'+a_4'}(a_5-a_3)\theta' \bigg\}} + E\bigg\{\frac{\theta'a_1'+a_2'}{\theta'a_3'+a_4'}(a_5-a_3) \theta\bigg\}\\
\end{align*}
Removing the redundant terms, we then have:
\begin{align*}
C'(\theta' - \theta) - T_2 + T_1
& = E\bigg\{ (\theta'-\theta) U_2 \frac{a_2'a_3'-a_1'a_4'}{(\theta a'_3 + a'_4)^2} \bigg\} + E\bigg\{\frac{\theta'a_1'+a_2'}{\theta'a_3'+a_4'}(a_5-a_3) \theta\bigg\}\\
&\quad + E\bigg\{ \frac{\theta'a_1' + a_2'}{\theta'a_3' + a_4'} (a_6 - a_4)
\bigg\} - E\bigg\{\frac{\theta a_1 + a_2}{\theta a_3 + a_4} \big[(a_5-a_3)\theta + a_6 - a_4\big]\bigg\} + T_1\\
&= E\bigg\{ \underbrace{(\theta'-\theta)U_2 \frac{a_2'a_3'-a_1'a_4'}{(\theta a'_3 + a'_4)^2}}_\text{$U_3$} \bigg\}
+ E\bigg\{\bigg[\underbrace{\frac{\theta'a_1'+a_2'}{\theta'a_3'+a_4'} - \frac{\theta a_1+a_2}{\theta a_3+a_4}}_\text{$U_4$}\bigg]\underbrace{[(a_5-a_3) \theta + a_6 - a_4]}_\text{$U_5$}\bigg\} + T_1
\end{align*}
We then have:
\begin{align*}
E(U_4U_5) &= \underbrace{E\bigg\{\frac{U_4U_5[(\theta' - \theta)a_3' + \theta(a_3' - a_3) + a_4' - a_4]}{\theta'a_3' + a_4'} \bigg\}}_\text{$R_1(\eta,\eta')$} + 
E\bigg\{\frac{U_4U_5(\theta a_3 + a_4)}{\theta'a_3' + a_4'} \bigg\} 
\end{align*}
Note that:
\begin{align*}
U_4
&= \frac{\theta'\theta a_1'a_3 + \theta'a_1'a_4 + \theta a_2'a_3 + a_2'a_4 - \theta\theta'a_1a_3' - \theta a_1a_4' - \theta'a_2a_3' - a_2a_4'}{(\theta'a_3'+a_4')(\theta a_3+a_4)}\\
&= \frac{\theta'\theta a_1'(a_3-a_3') + \theta'\theta a_3'(a_1'-a_1) + (\theta' - \theta)a_1'a_4 + \theta a_1'(a_4 - a_4') + \theta a_4'(a_1'-a_1)}{(\theta'a_3'+a_4')(\theta a_3+a_4)}\\
&\quad\quad + \frac{(\theta - \theta')a_2'a_3 + \theta' a_2'(a_3 - a_3') + \theta' a_3'(a_2'-a_2) + a_2'(a_4 - a_4') + a_4'(a_2' - a_2)}{(\theta'a_3'+a_4')(\theta a_3+a_4)}\\
& = \frac{\theta'(\theta a_1'+a_2')(a_3 - a_3') + \theta(a_1'-a_1)(\theta'a_3'+a_4')
+ (\theta' - \theta)(a_1'a_4-a_2'a_3) 
}
{(\theta'a_3'+a_4')(\theta a_3+a_4)}\\
& \quad\quad + \frac{ (a_4 - a_4')(\theta a_1' + a_2') + (a_2'-a_2)(\theta'a_3' + a_4')}
{(\theta'a_3'+a_4')(\theta a_3+a_4)}
\end{align*}
This hence implies that $R_1(\eta,\eta')$ is a second-order term. \\\\
Denote $M =   (\theta a_3 + a_4)[(\theta'a_3'+a_4')^2(\theta a_3+a_4)]^{-1}$, we also have:
\begin{align*}
R_2(\eta,\eta') &= E(U_3) + T_1 + E\bigg\{\frac{U_4U_5(\theta a_3 + a_4)}{\theta'a_3' + a_4'} \bigg\} \\
&= M \times \bigg\{\big[(\theta' - \theta)(a_2'a_3' - a_1'a_4') + \theta'(a_1-a_1')(\theta a_3' + a_4') + (\theta'a_2 + \theta'^2a_1')(a_3'-a_3) +\\
& \quad\quad\quad + (\theta'a_3+a_4)(a_2-a_2') + (a_2 + \theta'a_1')(a_4' - a_4)\big][\theta'(a_5 - a_3) + a_6 - a_4] \bigg\} +\\
& \quad M\times \bigg\{\big[ \theta'(\theta a_1' + a_2')(a_3 - a_3') + \theta(a_1' - a_1)(\theta'a_3'+a_4') + (\theta' - \theta)(a_1'a_4 - a_2'a_3)\big] + \\
& \quad\quad\quad + (a_4 - a_4')(\theta a_1' + a_2') + (a_2' - a_2)(\theta'a_3' + a_4') \big][\theta(a_5 - a_3) + a_6 - a_4]
\bigg\}
\end{align*}
It can be shown quite easily that this is a second-order term by grouping the products with a common factor. For instance, group the two products having $(\theta' - \theta)$, we have:
\begin{align*}
&(\theta' - \theta)(a_2'a_3' - a_1'a_4')[\theta'(a_5 - a_3) + a_6 - a_4] + (\theta' - \theta)(a_1'a_4 - a_2'a_3)[\theta(a_5 - a_3) + a_6 - a_4]\\
&= (\theta' - \theta)(a_5 - a_3)\big[a_2'(\theta'a_3'-\theta a_3) - a_1'(a_4'\theta'-a_4\theta) \big] + (\theta' - \theta)(a_6 - a_4)(a_2'a_3' - a_1'a_4' + a_1'a_4- a_2'a_3)\\
&= (\theta' - \theta)(a_5 - a_3)\big[a_2'(\theta'(a_3'-a_3) + a_3(\theta'-\theta)) - a_1'(a_4'(\theta'-\theta) + (a_4'-a_4)\theta) \big] \\
&+ (\theta' - \theta)(a_6 - a_4)(a_2'(a_3'-a_3) - a_1'(a_4' - a_4))
\end{align*}
The proof is similar for the other products in $R_2(\eta,\eta')$. As $R(\eta,\eta') = R_1(\eta,\eta') + R_2(\eta,\eta')$, we obtain the desired result.
\subsubsection{Influence functions when $\bm X$ are effect modifiers}
When $\beta = \beta(\bm X) = \beta_0 + \bm \beta_1^T\bm X$ where $\bm X$ is a vector of dimension $(K-1)$, the derivation from section 1 of this appendix is adjusted as follow:
\begin{align*} 
 \bm 0 &= E_\mathcal{P}\bigg[\bigg( \frac{\bm a_2a_3-\bm a_1a_4}{(\theta(\bm X) a_3 + a_4)^2}(\theta(\bm X)'_{t=0}) - \frac{\bm a'_{1,t=0}\theta(\bm X) + \bm a'_{2,t=0}}{\theta a_{3} + a_{4}} + \frac{(\theta(\bm X) \bm a_1 + \bm a_2)(a'_{3,t=0}\theta(\bm X) + a'_{4,t=0})}{(\theta(\bm X)  a_3 + a_4)^2}\bigg) B(\bm O) \bigg]\\
 &+ E_\mathcal{P}\bigg[\bm A(\bm O)(\theta'_{t=0}(\bm X))(Y_1D_1 - Y_0D_0)\bigg] +  \bm A(\tilde{\bm o},\mathcal{P}) B(\tilde{\bm o},\mathcal{P})
 \end{align*}
 where $\theta(\bm X) = e^{\bm \beta\bm X} - 1$, $\bm a_1(\bm X) = E\{Y_{0}D_{0}\bm d(\bm X,Z)|\bm X\}$ and $ \bm a_2(\bm X)  = E\{Y_0\bm d(\bm X,Z)|\bm X\}$, with $\bm d(\bm X, Z)$ being a $K$-dimensional vector function of $\bm X$ and $Z$.
 
 Note that $\theta(\bm X)'_{t = 0} = e^{\beta(\bm X)} X_*^{T}\bm\beta'_{t = 0}$, where $\bm X_* = \begin{pmatrix} 1 & \bm X^T\end{pmatrix}^T$ and $\bm\beta'_{t = 0} = \begin{pmatrix} \beta'_{0,t=0} & \beta'_{1,t=0}  \end{pmatrix}^T$. Plugging this into the above equation, one then has:
\begin{align*} 
 \bm 0 &= E_\mathcal{P}\bigg[\bigg( \frac{\bm a_2a_3-\bm a_1a_4}{(\theta(\bm X) a_3 + a_4)^2}e^{\beta(\bm X)} X_*^{T}\bm\beta'_{t = 0} - \frac{\bm a'_{1,t=0}\theta(\bm X) + \bm a'_{2,t=0}}{\theta a_{3} + a_{4}} + \frac{(\theta(\bm X) \bm a_1 + \bm a_2)(a'_{3,t=0}\theta(\bm X) + a'_{4,t=0})}{(\theta(\bm X)  a_3 + a_4)^2}\bigg) B(\bm O) \bigg]\\
 &+ E_\mathcal{P}\bigg[\bm A(\bm O)(e^{\beta(\bm X)} X_*^{T}\bm\beta'_{t = 0})(Y_1D_1 - Y_0D_0)\bigg] +  \bm A(\tilde{\bm o},\mathcal{P}) B(\tilde{\bm o},\mathcal{P})
 \end{align*} 
Hence,
 \begin{align*} 
&  E_\mathcal{P}\bigg[\bigg( \frac{\bm a_2a_3-\bm a_1a_4}{(\theta(\bm X) a_3 + a_4)^2}(e^{\beta(\bm X)} X_*^{T})B(\bm O) + \bm A(\bm O)(e^{\beta(\bm X)} X_*^{T})(Y_1D_1-Y_0D_0) \bigg)\bigg](\bm \beta'_{t=0}) \\
& = E_\mathcal{P}\bigg[\bigg(\frac{\bm a'_{1,t=0}\theta(\bm X) + \bm a'_{2,t=0}}{\theta(\bm X) a_{3} + a_{4}} - \frac{(\theta(\bm X) \bm a_1 + \bm a_2)(a'_{3,t=0}\theta(\bm X) + a'_{4,t=0})}{(\theta(\bm X) a_3 + a_4)^2}\bigg) B(\bm O) \bigg] -  \bm A(\tilde{\bm o},\mathcal{P}) B(\tilde{\bm o},\mathcal{P})
 \end{align*}
We thus have:
\[ 
\bm \beta'_{t=0} = \bm C^{-1} E_\mathcal{P}\bigg[\bigg(\frac{\bm a'_{1,t=0}\theta(\bm X) + \bm a'_{2,t=0}}{\theta(\bm X) a_{3} + a_{4}} - \frac{(\theta(\bm X) \bm a_1 + \bm a_2)(a'_{3,t=0}\theta(\bm X) + a'_{4,t=0})}{(\theta(\bm X) a_3 + a_4)^2}\bigg) B(\bm O) \bigg] -  \bm C^{-1}\bm A(\bm O) B(\bm O)
\]
where:
\[\bm C =  E_\mathcal{P}\bigg[\bigg( \frac{\bm a_2a_3-\bm a_1a_4}{(\theta(\bm X) a_3 + a_4)^2}(e^{\beta(\bm X)} X_*^{T})B(\bm O) + \bm A(\bm O)(e^{\beta(\bm X)} X_*^{T})(Y_1D_1-Y_0D_0) \bigg)\bigg]\]
and $\bm a'_{i,t=0}$ $(i=\overline{1,4})$ can be expressed in closed-form as in section 1 (with some adjustments to reflect the dimension of $\bm a_i(\bm X)$ and $\bm d(\bm X, Z)$).

Following the same reasoning as in section B.2.3, one can also show that 
\[\bm R(\bm \eta,\bm \eta') := \bm \beta(\bm \eta') - \bm \beta(\bm \eta) + E\{\bm\beta'_{t=0}(O,\bm\eta')\}\]
is a second order vector involving only products of the type $E[ \bm c(\bm \eta,\bm \eta_1)\{\bm f(\bm \eta')-\bm f(\bm \eta)\}\{\bm g(\bm \eta')-\bm g(\bm \eta)\}]$.

\end{document}